\documentclass[lettersize,journal]{IEEEtran}

\usepackage{amsmath,amsfonts}
\usepackage{algorithmic}
\usepackage{algorithm}
\usepackage{array}
\usepackage{subfig}
\usepackage{textcomp}
\usepackage{stfloats}
\usepackage{url}
\usepackage{graphicx}
\usepackage{cite}
\usepackage{booktabs}
\usepackage{multirow}
\usepackage{siunitx}
\usepackage{tikz}
\usepackage{orcidlink}
\usetikzlibrary{arrows.meta,positioning,calc,fit}
\hypersetup{hidelinks}

\begin{document}

\title{FAROS: Forecast-Aware Risk-Informed Online Reinforcement Learning for Proactive O-RAN Slicing}

\author{Peyman~Eftekhari\orcidlink{0009-0007-4198-392X},~\IEEEmembership{Student Member, IEEE}, Seyyed~Kasra~Mortazavi\orcidlink{0000-0001-9325-5130},~\IEEEmembership{Student Member, IEEE}, Reza~Farahani\orcidlink{0000-0002-2376-5802},~\IEEEmembership{Member, IEEE},\\ Robert~Legenstein\orcidlink{0000-0002-8724-5507}, Kyandoghere~Kyamakya\orcidlink{0000-0003-0773-9476},~\IEEEmembership{Member, IEEE}, and Martin~Gebser\orcidlink{0000-0002-8010-4752},~\IEEEmembership{Member, IEEE}%
\thanks{P. Eftekhari and M. Gebser are with the Institute of Artificial Intelligence and Cybersecurity, University of Klagenfurt, Klagenfurt, Austria (e-mails: peyman.eftekhari@aau.at; Martin.Gebser@aau.at).}%
\thanks{S. K. Mortazavi and K. Kyamakya are with the Institute of Smart Systems Technologies, University of Klagenfurt, Klagenfurt, Austria (e-mails: seyyedkasmor@edu.aau.at; kyandoghere.kyamakya@aau.at).}%
\thanks{R. Farahani is with the Institute of Information Systems Engineering, Distributed Systems Group, TU Wien, Vienna, Austria (e-mail: r.farahani@dsg.tuwien.ac.at).}%
\thanks{R. Legenstein is with the Institute of Machine Learning and Neural Computation, Graz University of Technology, Graz, Austria (e-mail: robert.legenstein@tugraz.at).}%
\thanks{P. Eftekhari and S. K. Mortazavi contributed equally to this work.}%
\thanks{This research was funded by the Austrian Science Fund (FWF) project 10.55776/COE12.}%
\thanks{Corresponding author: Reza Farahani (e-mail: r.farahani@dsg.tuwien.ac.at).}}

\maketitle

\begin{abstract}
Proactive Open Radio Access Network (O-RAN) slicing requires decisions before an approaching quality-of-service (QoS) loss becomes difficult to reverse. We present FAROS, a Forecast-Aware Risk-Informed O-RAN Slicing framework that combines an eight-step QoS forecast, training-only PCMCI+-supported temporal risk, Multi-Agent Proximal Policy Optimization (MAPPO) under centralized training and decentralized execution (CTDE), symbolic feasibility constraints, and reliability-aware Top-$K$ arbitration. MAPPO supplies the actor proposal, while A5v2 checks admissibility and one-step utility before execution. Evaluation uses ten validation-selected frozen checkpoints, 500 matched start indices per seed under the risky-start protocol, and 150,000 agent-level decisions. FAROS obtains mean reward $0.2778$ and ten-step recovery $90.56\%$ across ten seeds. A fixed model-matched diagnostic rule reaches reward $0.3015$ and recovery $97.57\%$ under the same transition--reward contract, so we make no superiority claim over direct threshold logic. A same-actor $M_4$ shield control also reaches zero risky-state prohibited actions. In the $M_5$ trace, A5v2 changes $22.61\%$ of actor proposals and reduces the predicate-defined prohibited-action rate from $16.79\%$ at proposal time to zero at execution; this property is enforced by the arbiter rather than learned by MAPPO. Independent ns-3 experiments test the six-action interface at packet level and remain separate from end-to-end FAROS claims.
\end{abstract}


\section{Introduction}
\label{sec:introduction}

O-RAN slicing can require intervention before service degradation becomes clear in current KPIs. Recent work uses O-RAN programmability and DRL for multi-agent control, VNF reconfiguration, slice orchestration, QoS-aware allocation, offloading, uncertainty, and SLA-aware management \cite{bonati2023openran_gym,amiri2024robust_vnf_oran,filali2024mvno_oran,dai2025sla_oran,cai2024online_slicing,mhatre2025qos_oran,liu2024ueqos_slicing,saleh2025eslice,sulaiman2023coordinated_slicing,chiang2023dql_slicing_offloading,gharehgoli2023ai_resource_slicing,fan2024mec_slicing}. Reactive control remains difficult when an approaching QoS loss is weak in the observed state.

Forecasting can expose this loss earlier. Mobile-network studies cover feature selection, spatial context, federated and direct prediction \cite{mendoza2023forecasting_mobile,zhou2023spatial_qos,phyu2023multislice_forecasting,perifanis2023federated_forecast}, and predictive information has been connected to policy transfer and proactive O-RAN orchestration \cite{nagib2023predictive_policy_transfer_ran_slicing,staffolani2024prorl}. Yet future QoS alone does not determine a command because the appropriate response also depends on resource state and vulnerability. The control problem is therefore to connect future QoS and temporal vulnerability with the learned multi-agent preference and execution-time admissibility. Constraint-aware RL and decentralized wireless control address parts of this problem \cite{yang2023distributional_safety_critic,hou2024interactive_safe_rl,du2023marl_dynamic_resource_6g,ji2024meta_federated_rl_resource_allocation,murti2024vran_reconfiguration,alsenwi2024embb_urllc}. FAROS addresses this execution path while retaining the learned actor.

FAROS combines current radio/resource measurements with an eight-step QoS forecast and PCMCI+-supported temporal risk in each cluster-level state. MAPPO learns the action distribution under centralized training and decentralized execution. PCMCI+ is restricted to training data; it supports the retained risk components and their adverse orientation without determining the deployed numerical weights.

The final $M_5$ stage applies A5v2 reliability-aware Top-$K$ arbitration. It retains strong actor proposals, filters candidates with frozen admissibility rules, and scores feasible actions using the evaluator's one-step transition and reward structure together with the frozen candidate-level action costs of Section~\ref{subsec:arbiter}. Actor probability and forecast reliability also enter the score; replacement occurs only when a feasible alternative exceeds the fixed intervention margin.

Evaluation preserves this separation. Leakage-safe forecasting and later-QoS vulnerability tests precede matched controller rollouts and proposal--execution audits. Sensitivity, runtime, calibration, and explainability characterize the frozen path, while independent ns-3 experiments check the semantics of the six commands outside end-to-end FAROS claims.

The main technical contributions are:
\begin{itemize}
    \item We formulate FAROS with three cluster-level actors under centralized training and decentralized execution. Each fixed local state combines current network measurements with an eight-step QoS forecast and temporal-risk information, while MAPPO retains decentralized actor proposals during execution.

    \item We integrate training-only PCMCI+ temporal discovery with the controller risk path while keeping it separate from action causality. The reconstructed structure supports the retained risk components and adverse orientation; numerical weights remain fixed implementation parameters. At $H=3$, adding temporal risk to the forecast raises recall from 0.083592 to 0.325462 and balanced accuracy from 0.540750 to 0.653257.

    \item We develop the final $M_5$ execution stage through the A5v2 reliability-aware Top-$K$ arbiter. It combines symbolic candidate admissibility, one-step model-based utility, actor preference, and forecast reliability. The audit follows all 150,000 agent-level proposal--execution records, separates actor behavior from enforced properties, and finds zero selected-to-executed mismatches.

    \item We evaluate FAROS on 500 matched start indices per training seed under the risky-start protocol. The same-actor $M_4$ shield control reduces the risky-state prohibited-action rate from 6.67\% to zero; $M_5$ has higher mean reward (0.27781 vs. 0.24649) and recovery (90.56\% vs. 80.87\%) under the same evaluator. A fixed model-matched diagnostic reference is reported separately; 120 matched ns-3 executions check the semantics of the six actions.
\end{itemize}

\section{Background and Related Work}
\label{sec:background}
\label{sec:related_work}

Learning-based O-RAN work covers allocation, VNF reconfiguration, slicing, SLA control, traffic steering, offloading, and service placement \cite{cai2024online_slicing,liu2024ueqos_slicing,mhatre2025qos_oran,amiri2024robust_vnf_oran,alsenwi2024embb_urllc,filali2024mvno_oran,dai2025sla_oran,kavehmadavani2024traffic_steering,sulaiman2023coordinated_slicing,ndikumana2023fed_dql_oran,monaco2026crown}. Forecasting spans spatial, multi-slice, federated, probabilistic, digital-twin, and edge settings \cite{mendoza2023forecasting_mobile,zhou2023spatial_qos,phyu2023multislice_forecasting,perifanis2023federated_forecast,dimauro2023voip_forecasting,kasuluru2025probabilistic_oran,navidan2026digital_twin_oran,tundo2025edge_forecasting}; predictive policy transfer and proactive orchestration connect forecasts to resource decisions \cite{nagib2023predictive_policy_transfer_ran_slicing,staffolani2024prorl}. These studies do not cover the path from early vulnerability to the executed command.

Multi-agent wireless control covers distributed resource management, service-function chaining, and RAN slicing \cite{du2023marl_dynamic_resource_6g,ji2024meta_federated_rl_resource_allocation,ai2025topology_marl,chai2025marl_micro_operator,birabwa2023marl_tntn}; safety-oriented work uses critics, constrained control, and shielding \cite{yang2023distributional_safety_critic,nagib2024safe_accelerated_oran_slicing,zangooei2024flexible_ran_slicing_cmarl,hsu2023shielding_generalization}. FAROS positions its execution-time intervention within this literature without treating predicate filtering as a safety certificate. PCMCI+ supplies training-only lagged-dependency support \cite{tian2024causal_effect_networks}, while the deployed risk score remains observational.

Table~\ref{tab:related_work_summary} gives the closest functional comparison; entries indicate reported functions rather than equivalent implementations. FAROS links future QoS and temporal risk to a learned multi-agent proposal, then applies Top-$K$ retention, admissibility filtering, forecast reliability, and margin-gated arbitration. This structure keeps forecasting, risk, proposals, and execution separately evaluable.

\begin{table*}[!t]
\caption{Positioning of FAROS Against Representative Recent Journal Studies.}
\label{tab:related_work_summary}
\centering
\scriptsize
\setlength{\tabcolsep}{3.0pt}
\renewcommand{\arraystretch}{1.08}
\begin{tabular}{p{2.65cm}cccccc}
\toprule
Work & \shortstack{O-RAN /\\slicing} & \shortstack{Predictive /\\forecast-aware} & \shortstack{Multi-agent\\RL} & \shortstack{Explicit\\constraints} & \shortstack{Reliability-aware\\arbitration} & \shortstack{Independent external /\\simulation evidence} \\
\midrule
Cai \emph{et al.} \cite{cai2024online_slicing} & Yes & -- & -- & -- & -- & -- \\
Nagib \emph{et al.} \cite{nagib2023predictive_policy_transfer_ran_slicing} & Yes & Yes & -- & -- & -- & -- \\
Nagib \emph{et al.} \cite{nagib2024safe_accelerated_oran_slicing} & Yes & Yes & -- & Yes & -- & -- \\
Zangooei \emph{et al.} \cite{zangooei2024flexible_ran_slicing_cmarl} & Yes & -- & Yes & Yes & -- & -- \\
Staffolani \emph{et al.} \cite{staffolani2024prorl} & Yes & Yes & -- & -- & -- & -- \\
Kavehmadavani \emph{et al.} \cite{kavehmadavani2024traffic_steering} & Yes & Yes & Yes & -- & -- & -- \\
Kasuluru \emph{et al.} \cite{kasuluru2025probabilistic_oran} & Yes & Yes & -- & -- & -- & -- \\
Saleh \emph{et al.} \cite{saleh2025eslice} & Yes & -- & -- & Yes & -- & -- \\
Monaco \emph{et al.} \cite{monaco2026crown} & Yes & -- & -- & Yes & -- & -- \\
\textbf{FAROS} & Yes & Yes & Yes & Yes & Yes & Yes \\
\bottomrule
\end{tabular}
\end{table*}

\section{System Architecture and Control Setting}
\label{sec:system_model}

\begin{figure*}[!t]
\centering
\begin{tikzpicture}[
    x=1mm,
    y=1mm,
    font=\footnotesize,
    >=Latex,
    main/.style={
        draw=black!88,
        fill=white,
        rounded corners=1.1pt,
        line width=0.72pt,
        align=center,
        text width=21.8mm,
        minimum height=11.4mm,
        inner sep=2.0pt,
        font=\footnotesize\bfseries
    },
    support/.style={
        draw=black!78,
        fill=white,
        rounded corners=1.0pt,
        line width=0.64pt,
        align=center,
        text width=29.0mm,
        minimum height=9.4mm,
        inner sep=1.8pt,
        font=\footnotesize\bfseries
    },
    flow/.style={
        -{Latex[length=1.9mm,width=1.15mm]},
        draw=black!88,
        line width=0.72pt,
        shorten >=1.6pt,
        shorten <=1.6pt
    },
    supportflow/.style={
        -{Latex[length=1.7mm,width=1.05mm]},
        draw=black!68,
        line width=0.64pt,
        shorten >=1.4pt,
        shorten <=1.4pt
    },
    heading/.style={
        font=\small\bfseries,
        text=black!82
    }
]

\node[heading] at (0,11.6) {Online FAROS Decision Path};

\node[main] (env)      at (-78,0) {O-RAN Slicing\\Environment};
\node[main] (forecast) at (-52,0) {Forecast\\Module};
\node[main] (risk)     at (-26,0) {Temporal Risk\\Module};
\node[main] (state)    at (  0,0) {Local State\\Assembly};
\node[main] (mappo)    at ( 26,0) {CTDE\\MAPPO};
\node[main] (exec)     at ( 52,0) {$M_5$/A5v2\\Execution};
\node[main] (action)   at ( 78,0) {Six-Action\\Interface};

\draw[flow] (env.east) -- (forecast.west);
\draw[flow] (forecast.east) -- (risk.west);
\draw[flow] (risk.east) -- (state.west);
\draw[flow] (state.east) -- (mappo.west);
\draw[flow] (mappo.east) -- (exec.west);
\draw[flow] (exec.east) -- (action.west);

\coordinate (fbRight) at (92,0);
\coordinate (fbTop)   at (92,7.2);
\coordinate (fbLeft)  at (-92,7.2);
\draw[flow]
    (action.east) -- (fbRight) -- (fbTop)
    -- (fbLeft) -- (-92,0) -- (env.west);

\node[heading] at (0,-17.0) {Training-Only Temporal Support};

\node[support] (train)      at (-54,-28.0) {Training Reservation\\Trajectories};
\node[support] (pcmci)      at (-18,-28.0) {PCMCI+ Temporal\\Discovery};
\node[support] (components) at ( 18,-28.0) {Supported Risk\\Components};
\node[support] (fixed)      at ( 54,-28.0) {Fixed Risk\\Construction};

\draw[supportflow] (train.east) -- (pcmci.west);
\draw[supportflow] (pcmci.east) -- (components.west);
\draw[supportflow] (components.east) -- (fixed.west);

\coordinate (supportUp) at (54,-11.5);
\coordinate (supportAcross) at (-26,-11.5);
\draw[supportflow]
    (fixed.north) -- (supportUp)
    -- (supportAcross)
    -- (risk.south);

\end{tikzpicture}
\caption{FAROS controller architecture with the online decision path and training-only temporal-discovery support path.}
\label{fig:oran_architecture}
\end{figure*}
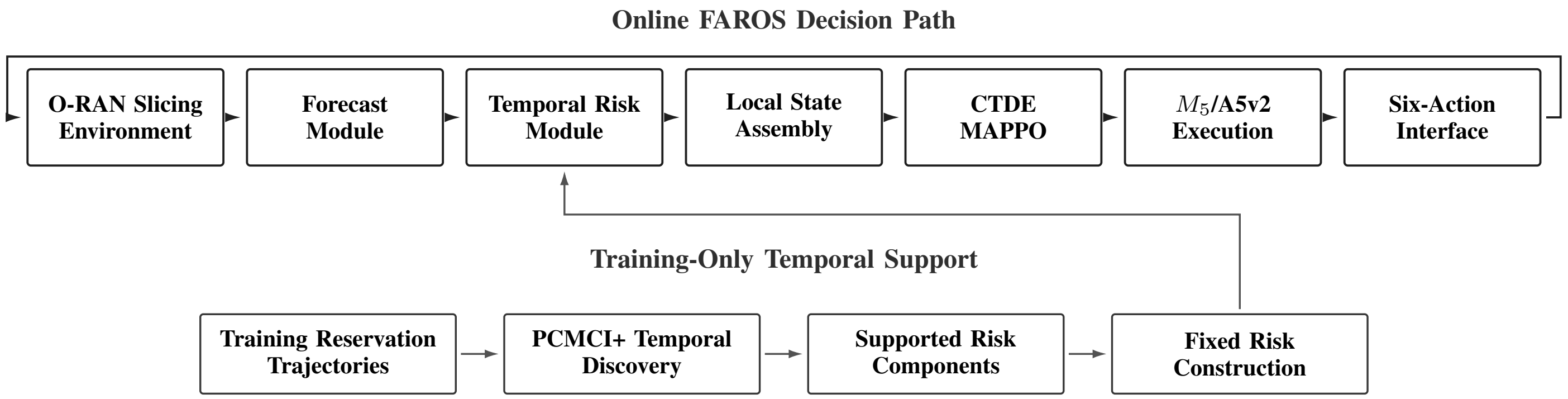

Fig.~\ref{fig:oran_architecture} separates the online FAROS path from training-only temporal discovery. It is a logical controller architecture rather than an implemented O-RAN interface stack. A prospective near-real-time RIC xApp can host forecasting, runtime risk, decentralized actors, shielding, and A5v2, while PCMCI+ discovery and critic optimization remain offline. The present implementation does not instantiate E2SM-KPM, E2SM-RC, A1, or O1 signaling or measure RIC/transport latency.

\subsection{O-RAN Slicing Environment}
\label{subsec:oran_environment}

Let $\mathcal{C}=\{c_1,\ldots,c_{N_c}\}$ denote the traffic clusters, with $N_c=3$ in evaluation. Each cluster executes a decentralized actor over the common six-action space, while the centralized critic uses the concatenated cluster states during training. The environment averages cluster-level rewards into a shared team reward. Its trace-conditioned transition model updates each cluster resource state separately and does not impose an explicit shared physical resource block (PRB) budget or inter-cluster contention constraint. The 10-dimensional local state combines current measurements with an eight-step LSTM forecast from a 64-step QoS history and training-only PCMCI+-supported risk. The actor proposal is executed directly, shielded, or reconsidered by $M_5$/A5v2. Paired evaluations keep the transition model, seeds, and matched starts fixed; ns-3 later checks the same action semantics.

PCMCI+ remains outside online action selection. Table~\ref{tab:action_space} defines the common commands used by every controller variant and by the ns-3 implementation.

\begin{table}[!t]
\caption{Six-Action Control Space.}
\label{tab:action_space}
\centering
\small
\begin{tabular}{cl}
\toprule
Action & Control command \\
\midrule
$A_0$ & Keep current configuration \\
$A_1$ & Increase resource fraction by 10\% \\
$A_2$ & Decrease resource fraction by 10\% \\
$A_3$ & Switch scheduler \\
$A_4$ & Switch scheduler and increase resources by 10\% \\
$A_5$ & Switch scheduler and decrease resources by 10\% \\
\bottomrule
\end{tabular}
\end{table}

\subsection{Multi-Agent Control Architecture}
\label{subsec:multiagent_architecture}

Under CTDE, the centralized critic receives concatenated cluster states during optimization, while each actor proposes from its local state at execution; agents share the mean cluster reward. The evaluator has no explicit shared PRB budget or cross-cluster contention, so no advantage over independent PPO is isolated. In $M_5$, A5v2 may reconsider the learned proposal before execution.

\section{Model and Problem Formulation}
\label{sec:problem_formulation}

This section formalizes the state, forecast, temporal-risk, MAPPO, and execution components of FAROS. Table~\ref{tab:notation} collects the symbols used below.

\begin{table*}[!t]
\caption{Main Notation Used in the FAROS Formulation.}
\label{tab:notation}
\centering
\scriptsize
\setlength{\tabcolsep}{4pt}
\renewcommand{\arraystretch}{1.12}
\begin{tabular}{p{2.15cm} p{5.55cm} p{2.15cm} p{5.55cm}}
\toprule
Symbol & Meaning & Symbol & Meaning \\
\midrule
$i$, $t$ &
Agent index and decision step &
$\mathcal{N}$, $N=|\mathcal{N}|$ &
Set of agents and number of agents \\

$\mathbf{s}_{i,t}$ &
Local state of agent $i$ at step $t$ &
$\mathbf{s}_t$ &
Joint state of all agents \\

$a_{i,t}$, $\mathbf{a}_t$ &
Local action and joint action &
$\mathcal{A}$ &
Common six-action set $\{A_0,\ldots,A_5\}$ \\

$\mathcal{S}$ &
Joint state space &
$P$, $R$, $\gamma_{\mathrm{disc}}$ &
Transition mechanism, reward function, and discount factor \\

$q_t$ &
Trace-derived QoS utility at step $t$ &
$\widehat{q}^{\,\mathrm{future}}_t$ &
Predicted future-QoS term used by the controller \\

$\widehat{\Delta q}_t$ &
Candidate-adjusted future QoS minus current QoS &
$\rho^{\,\mathrm{causal}}_t$ &
Temporal-risk score supplied to the controller \\

$\mathbf{h}_t$ &
64-step QoS history used by the forecaster &
$\widehat{q}_{t+8}$ &
Eight-step-ahead QoS forecast target \\

$f_{\theta_f}$, $\theta_f$ &
LSTM forecasting function and trained parameters &
$\mathbf{z}_t$ &
Variable vector used for PCMCI+ temporal discovery \\

$\mathcal{E}_q$ &
FDR-controlled lagged-link set into QoS &
$\tau$ &
Temporal lag used in PCMCI+ discovery \\

$\pi_{\theta_i}$ &
Actor policy of agent $i$ &
$V_{\phi}$ &
Centralized critic with parameters $\phi$ \\

$r_t(\theta_i)$ &
PPO probability ratio for agent $i$ &
$\widehat{A}_{i,t}$ &
Normalized advantage estimate \\

$\tau_q^{(M_3)}$, $\tau_\rho^{(M_3)}$ &
Frozen low-QoS and high-risk thresholds of the $M_3$ shield &
$\tau_q^{(M_5)}$, $\tau_\rho^{(M_5)}$ &
Frozen low-QoS and high-risk thresholds of the $M_5$ arbiter \\

$a^{\mathrm{actor}}_{i,t}$ &
Actor argmax proposal &
$\mathcal{C}_{i,t}$ &
Top-$K$ candidate set retained by A5v2 \\

$z_{i,t}$ &
Risky-state indicator used by $M_5$ &
$\mathcal{C}^{\mathrm{feas}}_{i,t}$ &
Feasible subset after symbolic filtering \\

$r^{\mathrm{rel}}_{i,t}$ &
Forecast-reliability score &
$\widehat{R}_{i,t}(a)$ &
One-step model-based utility of candidate action $a$ \\

$P_{i,t}(a)$ &
Actor-preference term used by A5v2 &
$G_{i,t}(a)$ &
Forecast-reliability adjustment for candidate $a$ \\

$S_{i,t}(a)$ &
Final A5v2 candidate score &
$w_{\mathrm{actor}}$, $w_{\mathrm{rel}}$ &
Frozen actor-preference and reliability weights \\

$\delta_{\mathrm{int}}$ &
Intervention margin &
$a^\star_{i,t}$ &
Highest-scoring feasible candidate selected by the arbiter \\

$b^{\mathrm{DL}}$, $\eta^{\mathrm{UL}}$, $p^{\mathrm{req}}$, $p^{\mathrm{grant}}$ &
DL bitrate, UL SINR, requested PRBs, and granted PRBs &
$n^{\mathrm{UE}}$, $\mu^{\mathrm{DL}}$, $u^{\mathrm{buf}}$ &
UE count, DL MCS, and UL-buffer occupancy \\

$C_{\mathrm{res}}$, $C_{\mathrm{sched}}$ &
Resource-reconfiguration and scheduler-switching costs &
$L$, $d_f$, $h_f$, $K$ &
Forecast look-back, forecast input dimension, LSTM hidden width, and retained candidate count \\
\bottomrule
\end{tabular}
\end{table*}

\subsection{Multi-Agent Decision Variables}
\label{subsec:multiagent}

At decision step $t$, agent $i$ corresponds to traffic cluster $c_i$,
receives a local state $\mathbf{s}_{i,t}\in\mathbb{R}^{10}$, and selects
$a_{i,t}\in\mathcal{A}$, where
\begin{equation*}
\mathcal{A}=\{A_0,A_1,A_2,A_3,A_4,A_5\}.
\end{equation*}
For compact notation, the collections of local states and actions are written as
\begin{equation*}
\mathbf{s}_t=
\left(\mathbf{s}_{1,t},\mathbf{s}_{2,t},\mathbf{s}_{3,t}\right),
\qquad
\mathbf{a}_t=
\left(a_{1,t},a_{2,t},a_{3,t}\right).
\end{equation*}
These joint symbols collect the three cluster variables used by the CTDE formulation; they do not imply a coupled resource transition or a shared PRB-capacity constraint.

\subsection{State and Action Representation}
\label{subsec:state_action}

All controller variants use the same normalized 10-dimensional local state,
\begin{equation*}
\small
\mathbf{s}_{i,t}=\big[
 b^{\mathrm{DL}}_{i,t},\eta^{\mathrm{UL}}_{i,t},p^{\mathrm{req}}_{i,t},p^{\mathrm{grant}}_{i,t},n^{\mathrm{UE}}_{i,t},
\mu^{\mathrm{DL}}_{i,t},u^{\mathrm{buf}}_{i,t},q_{i,t},
\widehat{q}^{\,\mathrm{future}}_{i,t},\rho^{\,\mathrm{causal}}_{i,t}\big]^\top.
\end{equation*}
This is the frozen feature order used by the actor: downlink bitrate, uplink SINR, requested and granted PRBs, UE count, downlink MCS, uplink-buffer occupancy, current QoS, predicted future QoS, and temporal risk. The order is unchanged across paired experiments, and no feature is added during final evaluation. Table~\ref{tab:action_space} keeps resource changes distinct from scheduler switching, which is necessary for interpreting $M_5$-versus-rule replays when similar mean resource levels arise from different action sequences.

\subsection{Markov Decision Process Formulation}
\label{subsec:mdp}

The control process is $\langle\mathcal{N},\mathcal{S},\mathcal{A},P,R,\gamma_{\mathrm{disc}}\rangle$, with three agents, the common six-action set, trace-conditioned transition mechanism, reward, and discount factor.

For the forecast- and temporal-risk-aware controllers, the controller reward is
\begin{equation}
r_t
=
1.0\,\widehat{q}^{\,\mathrm{future}}_t
+
0.5\,\widehat{\Delta q}_t
-
0.1\,\rho^{\,\mathrm{causal}}_t,
\label{eq:frozen_reward}
\end{equation}
where $\widehat{q}^{\,\mathrm{future}}_t$ and $\rho^{\,\mathrm{causal}}_t$ denote predicted future QoS and temporal risk, respectively, and $\widehat{\Delta q}_t=\widehat{q}^{\,\mathrm{future,adj}}_t-q_t$ is the candidate-adjusted future QoS relative to the current QoS. Equation~\eqref{eq:frozen_reward} favors predicted QoS and improvement while penalizing risk. The versioned implementation fixes the coefficients at $1.0$, $0.5$, and $0.1$; they are protocol parameters rather than optimized weights.

Table~\ref{tab:transition_coeffs} reports the frozen QoS and risk response coefficients of the trace-conditioned evaluator. They are modeling assumptions recovered from the implementation, rather than measured radio parameters; Section~\ref{subsec:transition_sensitivity} tests their influence.

\begin{table}[!t]
\caption{Transition-Model Coefficients.}
\label{tab:transition_coeffs}
\centering
\small
\begin{tabular}{lr}
\toprule
Coefficient & Value \\
\midrule
\texttt{qos\_prb\_coeff} & 0.08 \\
\texttt{risk\_prb\_coeff} & -0.20 \\
\texttt{risky\_idle\_qos\_penalty} & 0.08 \\
\texttt{risky\_idle\_risk\_penalty} & 0.20 \\
\texttt{risky\_safe\_qos\_bonus} & 0.10 \\
\texttt{risky\_safe\_risk\_bonus} & -0.15 \\
\texttt{risky\_reduce\_qos\_penalty} & 0.25 \\
\texttt{risky\_reduce\_risk\_penalty} & 0.60 \\
\bottomrule
\end{tabular}
\end{table}

Controller-level gains are conditional on this transition contract. We therefore vary actuator strength while holding the controller fixed, and use ns-3 separately to test the semantics of the six actions outside the trace-conditioned evaluator.

The QoS variable $q_t$ used by the forecast, temporal-risk, and controller paths is the trace-derived composite utility retained in the controller data. For each constituent $x_t$, let $\widetilde{x}_t=(x_t-x_{\min})/(x_{\max}-x_{\min}+10^{-8})$. The resulting utility is
\begin{equation}
q_t =
0.45\widetilde{T}^{DL}_t
+0.35\widetilde{G}^{PRB}_t
-0.15\widetilde{R}^{PRB}_t
-0.05\widetilde{N}^{UE}_t ,
\end{equation}
where the terms denote normalized downlink throughput, granted PRBs, requested PRBs, and UE load, respectively. Higher throughput and granted resources increase $q_t$, whereas greater requested resources and UE load decrease it. We use this trace-derived utility consistently as the QoS target in the forecasting and controller evaluations.

\subsection{Forecasting Formulation}
\label{subsec:forecast_module}

The first information path gives the controller a view of near-future QoS before an action is chosen, following recent mobile-network forecasting studies that include recurrent and LSTM-based models \cite{mendoza2023forecasting_mobile,zhou2023spatial_qos,dimauro2023voip_forecasting}. The LSTM uses the 64-step history available at decision step $t$,
\begin{equation*}
\mathbf{h}_{t}=[q_{t-63},q_{t-62},\ldots,q_t],
\end{equation*}
to predict QoS eight steps ahead,
\begin{equation*}
\widehat{q}_{t+8}=f_{\theta_f}(\mathbf{h}_t),
\end{equation*}
where $\theta_f$ denotes the forecasting parameters. The retained controller data establish the 64-step input history, the eight-step target, and the stored $\widehat{q}^{\,\mathrm{future}}_t$ values used by FAROS. The original LSTM training source and complete optimization recipe are not preserved, so full-trace statistics from that stored forecast are not interpreted as holdout accuracy. This forecast horizon is separate from the controller-evaluation horizon.

Forecast capability is therefore checked independently with a leakage-safe LSTM trained from the retained QoS traces under a frozen chronological protocol. Within each eligible cluster--reservation trace, 64-step inputs and their $t+8$ targets are assigned to train, validation, and test partitions without target overlap. Scaling is fitted on training data only, checkpoint selection uses validation MAE only, and the protected test partition is evaluated once after selection. This validation model is used only to test the forecasting capability; it does not replace the stored forecast values consumed by the frozen FAROS experiments. Section~\ref{subsec:forecast_results} reports MAE, RMSE, $R^2$, and Pearson correlation against a past-only eight-sample moving average (MovingAverage8) and persistence, and separately tests evaluation-time forecast sensitivity.

\subsection{Temporal-Risk Formulation}
\label{subsec:causal_module}

The second information path represents temporal vulnerability from lagged dependencies among the radio and traffic variables. The notation $\rho_t^{\mathrm{causal}}$ follows the controller implementation; throughout the paper it denotes the PCMCI+-supported temporal-risk score and is not interpreted as an action-level causal effect. We reconstruct these dependencies with PCMCI+ through Tigramite, following recent journal use of PCMCI+ for time-series dependency analysis \cite{tian2024causal_effect_networks}, and use partial correlation (ParCorr) as the conditional-independence test. The discovery set is
\begin{equation*}
\mathbf{z}_t =
\left[
n_t^{\mathrm{UE}},
p_t^{\mathrm{req}},
p_t^{\mathrm{grant}},
b_t^{\mathrm{DL}},
\eta_t^{\mathrm{UL}},
\mu_t^{\mathrm{DL}},
u_t^{\mathrm{buf}},
q_t
\right]^\top,
\end{equation*}
where the entries correspond to \texttt{num\_ues},
\texttt{sum\_requested\_prbs}, \texttt{sum\_granted\_prbs},
\texttt{tx\_brate\_downlink\_mbps}, \texttt{ul\_sinr},
\texttt{dl\_mcs}, \texttt{ul\_buffer\_bytes}, and
\texttt{qos\_score}, respectively.

The reconstruction uses only the training region and preserves reservation boundaries. Independent reservation trajectories are passed to Tigramite in multiple-series mode, preventing lagged links across reservations. The conditional-independence test is evaluated for
\begin{equation*}
\tau \in \{1,\ldots,8\},
\end{equation*}
with $\tau_{\min}=1$, $\tau_{\max}=8$, and
$\alpha_{\mathrm{PC}}=0.05$. Benjamini--Hochberg false-discovery-rate correction is applied at 0.05, consistent with multiplicity-controlled PCMCI analysis \cite{tian2024causal_effect_networks}. Temporal-dependency discovery uses the 518,430 training observations (172,810 per cluster) organized as 90 independent reservation trajectories. Validation and final-test observations are excluded.

For a candidate lagged relation
$z_{j,t-\tau}\rightarrow q_t$, PCMCI+ returns a partial-correlation coefficient
$\beta_{j,\tau}$ and an associated significance value. We denote the retained
FDR-controlled set by
\begin{equation*}
\mathcal{E}_q =
\left\{
(j,\tau):
p^{\mathrm{FDR}}_{j,\tau}<0.05
\right\}.
\end{equation*}
The reconstruction identifies 20 significant lagged links into the QoS target. In
particular, both variables used by the historical risk index,
\texttt{sum\_granted\_prbs} and \texttt{dl\_mcs}, retain significant lagged links to
QoS. Granted PRBs are significant at lags $\{2,5,6,8\}$, while downlink MCS is
significant at lags $\{2,3,4\}$. Their aggregate signed partial correlations are
negative, consistent with the adverse orientation used in the risk construction.

The controller does not recompute a graph at every decision. It consumes a scalar
temporal-risk index constructed from the two historical PCMCI+-supported components. Let
\begin{equation*}
g_t=-p_t^{\mathrm{grant}},
\qquad
d_t=-\mu_t^{\mathrm{DL}},
\end{equation*}
and standardize each component as
\begin{equation*}
\bar{x}_t =
\frac{x_t-\mu_x}{\sigma_x+10^{-8}},
\qquad
x\in\{g,d\}.
\end{equation*}
The unscaled risk index is
\begin{equation}
r_t^{(0)}
=
0.709609\,\bar{g}_t
+
0.290391\,\bar{d}_t.
\label{eq:causal_risk_unscaled}
\end{equation}
A final standardization gives the temporal-risk score supplied to the controller,
\begin{equation*}
\rho_t^{\mathrm{causal}}
=
\frac{
r_t^{(0)}-\mu_{r^{(0)}}
}{
\sigma_{r^{(0)}}+10^{-8}
}.
\end{equation*}

The weights in \eqref{eq:causal_risk_unscaled} are the frozen historical parameters used by the controller. The preserved implementation records this pair directly in the PCMCI+-supported risk construction. We therefore retain the exact operational weights without interpreting them as raw PCMCI+ partial-correlation coefficients. The training-only reconstruction is used to verify the retained components, their lagged links to QoS, and their adverse orientation; Table~\ref{tab:pcmci_config} records this discovery contract separately from reward and execution logic.

\begin{table*}[!t]
\caption{PCMCI+ Temporal-Dependency Discovery Configuration.}
\label{tab:pcmci_config}
\centering
\small
\begin{tabular}{ll}
\toprule
Property & Configuration \\
\midrule
Conditional-independence test & ParCorr \\
Discovery variables & UE count, requested PRBs, granted PRBs, DL bitrate, UL SINR, DL MCS, UL buffer, QoS \\
Target variable & \texttt{qos\_score} \\
Lag interval & $\tau=1,\ldots,8$ \\
PC significance level & $0.05$ \\
Multiplicity correction & Benjamini--Hochberg FDR, $\alpha=0.05$ \\
Discovery-fit observations & 518,430 \\
Independent reservation trajectories & 90 \\
Cross-reservation lagged links & Disallowed \\
Significant lagged links into QoS & 20 \\
Risk components retained by controller & \texttt{sum\_granted\_prbs}, \texttt{dl\_mcs} \\
Historical component weights & 0.709609, 0.290391 \\
\bottomrule
\end{tabular}
\end{table*}

For consistency with the retained controller data, the descriptive risk analysis uses the 1,001,456 aligned controller observations. Across numerical quartiles of the temporal-risk score, mean QoS decreases from 0.264621 in Q1 to $-0.002944$ in Q4, with Pearson correlation $r=-0.9687$. These results are associative; future vulnerability is tested separately in Section~\ref{subsec:causal_results}.

\subsection{MAPPO Optimization Model}
\label{subsec:mappo_controller}

MAPPO is the learned policy component of FAROS and converts the assembled local state into action preferences under centralized training and decentralized execution, following recent journal formulations of MAPPO and multi-agent resource control \cite{hassan2024comappo_tmc,du2023marl_dynamic_resource_6g,zangooei2024flexible_ran_slicing_cmarl}. Each actor maps $10\!\to\!128\!\to\!128\!\to\!6$ with Tanh activations, while the centralized critic maps the concatenated 30-dimensional state through two 256-unit Tanh layers to one value. FAROS adds forecast and temporal-risk inputs to this policy path and governs the actor proposal through shielding or A5v2 arbitration. Since the experiments do not compare MAPPO with independent PPO, we do not attribute controller gains specifically to centralized training.

For a policy update, we use the clipped PPO formulation described in \cite{zheng2025rewardppo}. The probability ratio for sample $t$ is
\begin{equation*}
r_t(\theta_i)
=
\frac{
\pi_{\theta_i}(a_{i,t}\mid\mathbf{s}_{i,t})
}{
\pi_{\theta_i^{\mathrm{old}}}(a_{i,t}\mid\mathbf{s}_{i,t})
}.
\end{equation*}
The clipped actor objective is
\begin{equation*}
\begin{array}{@{}l@{}}
L_i^{\mathrm{clip}}(\theta_i)
=
{} \\ \qquad\quad
\mathbb{E}_t
\left[
\min
\left(
r_t(\theta_i)\widehat{A}_{i,t},
\operatorname{clip}(r_t(\theta_i),1-\epsilon,1+\epsilon)
\widehat{A}_{i,t}
\right)
\right],
\end{array}
\end{equation*}
where $\widehat{A}_{i,t}$ is the normalized advantage estimate. The retained MAPPO implementation uses generalized advantage estimation in the standard PPO-based multi-agent optimization role \cite{hassan2024comappo_tmc,zheng2025rewardppo}, with $\gamma_{\mathrm{disc}}=0.99$ and $\lambda_{\mathrm{GAE}}=0.95$; the PPO clipping parameter is $\epsilon=0.20$. The critic minimizes
\begin{equation*}
L^{V}(\phi)
=
\mathbb{E}_t
\left[
\left(V_{\phi}(\mathbf{s}_t)-\widehat{R}_t\right)^2
\right].
\end{equation*}
Actor and critic use Adam at $3\times10^{-4}$ with value-loss coefficient 0.5 and entropy coefficient 0.005. Training uses 512-transition rollouts, five PPO epochs, and 60 iterations, with validation every five iterations on 128 protected starts. For each seed from 0 to 9, checkpoint selection uses the maximum validation score, with the earliest iteration resolving a tie. The ten retained checkpoints have distinct verified hashes and share the same training-configuration and validation-index fingerprints. Protected final-test outcomes are excluded from checkpoint selection and are accessed only after the checkpoints are frozen.

Table~\ref{tab:method_lineage} separates the FAROS ablation ladder from the independent Fair MAPPO reference and from the physical actions. $M_0$ is the no-forecast/no-risk ablation within $M_0$--$M_5$; it is not Fair MAPPO. Fair MAPPO is the separately trained \texttt{Standard\_MAPPO\_Common\_Dynamics\_V2} baseline: forecast and risk inputs are masked from actor and critic, the common dynamics and six-action space are retained, and the shield is disabled. Its ten seed-specific policies use 128-unit actors and a 256-unit centralized critic and are trained for 30 iterations of 512 rollout steps with Adam $3\times10^{-4}$, PPO clipping 0.20, entropy coefficient 0.02, value coefficient 0.5, and five PPO epochs. The frozen seed-specific checkpoints are evaluated without retraining on the same 500 matched start indices selected under the risky-start protocol and the same 10-step horizon, yielding 150,000 decisions across ten seeds. $A_0$--$A_5$ remain the common physical commands.

\begin{table}[!t]
\caption{Controller Lineage.}
\label{tab:method_lineage}
\centering
\small
\begin{tabular}{p{1.55cm}p{6.65cm}}
\toprule
Method & Configuration \\
\midrule
$M_0$ & No-Forecast/No-Risk MAPPO \\
$M_1$ & Forecast-Only MAPPO \\
$M_2$ & Temporal-Risk-Only MAPPO \\
$M_3$ & Shielded Forecast+Temporal-Risk MAPPO \\
$M_4$ & Forecast+Temporal-Risk MAPPO; actor prior of $M_5$ \\
$M_4$+Shield & $M_4$ actor + $M_5$ risky-state $A_2/A_5$ prohibition only \\
$M_5$ & FAROS reliability-aware Top-$K$ arbiter over $M_4$ \\
Fair MAPPO & Independent common-dynamics MAPPO; forecast/risk masked \\
\bottomrule
\end{tabular}
\end{table}

\subsection{Execution Constraints}
\label{subsec:shield}

Following the execution-constraint perspective used in safe and constrained RL \cite{yang2023distributional_safety_critic,hou2024interactive_safe_rl,nagib2024safe_accelerated_oran_slicing,zangooei2024flexible_ran_slicing_cmarl}, $M_3$ adds a state-dependent shield to the forecast- and temporal-risk-aware policy. In the reconstructed $M_3$ configuration, the risky-state thresholds were estimated from 518,430 training-only observations. The low-QoS boundary corresponds to the 0.35 quantile,
\begin{equation}
\tau_q^{(M_3)} \approx 0.079330,
\end{equation}
and the high-risk boundary corresponds to the 0.65 quantile,
\begin{equation}
\tau_\rho^{(M_3)} \approx 0.198715.
\end{equation}
These reconstructed $M_3$ thresholds remain separate from $M_5$/A5v2; displayed values are rounded, while evaluation retains machine precision.

The final $M_5$ stage follows the frozen A5v2 arbitration configuration used before the protected final evaluation. Its low-QoS and high-risk boundaries are
\begin{equation}
\tau_q^{(M_5)} \approx 0.064818,
\qquad
\tau_\rho^{(M_5)} \approx 0.442658.
\label{eq:m5_thresholds}
\end{equation}
A5v2 computes these boundaries as the 0.35 predicted-future-QoS and 0.65 temporal-risk quantiles of its reference controller dataframe, whose scope differs from the training-only $M_3$ data. Evaluation retains machine precision; protected final-test outcomes do not estimate or tune either pair.

For evaluation, a \emph{predicate-defined prohibited action} is $A_2$ or $A_5$ while the frozen risky-state predicate is active. A zero executed rate therefore records symbolic enforcement, rather than learned MAPPO safety or formal network safety. Reward, recovery, QoS and risk changes, resource use, and ns-3 assess operational behavior separately.

\subsection{Reliability-Aware Arbitration}
\label{subsec:arbiter}

$M_5$ uses the A5v2 arbitration procedure after the $M_4$ Forecast+Temporal-Risk MAPPO actor has produced a probability
distribution over the six actions. Let
$\pi_{\theta_i}(a\mid\mathbf{s}_{i,t})$ denote the actor probability for agent $i$.
The actor proposal is
\begin{equation*}
a^{\mathrm{actor}}_{i,t}
=
\arg\max_{a\in\mathcal{A}}
\pi_{\theta_i}(a\mid\mathbf{s}_{i,t}),
\end{equation*}
and the arbiter retains the $K$ actions with the largest actor probabilities,
\begin{equation*}
\mathcal{C}_{i,t}
=
\operatorname{TopK}_{a\in\mathcal{A}}
\pi_{\theta_i}(a\mid\mathbf{s}_{i,t}),
\qquad K=3.
\end{equation*}
The actor argmax is explicitly retained in this candidate set.

The risky-state indicator for the final $M_5$ controller uses the frozen A5v2 thresholds in \eqref{eq:m5_thresholds},
\begin{equation*}
z_{i,t}
=
\mathbb{I}\!\left[
\widehat{q}^{\,\mathrm{future}}_{i,t}<\tau_q^{(M_5)}
\;\lor\;
\rho^{\,\mathrm{causal}}_{i,t}>\tau_\rho^{(M_5)}
\right].
\end{equation*}
Before candidate scoring, A5v2 applies symbolic admissibility rules. When $z_{i,t}=1$, the PRB-reduction actions $A_2$ and $A_5$ are inadmissible. The idle action $A_0$ is also inadmissible when the forecast-reliability score $r^{\mathrm{rel}}_{i,t}$ is below $0.42$. Reliability is computed from forecast errors that have already matured at decision time. With forecast horizon $H=8$ and reliability window $W=64$, the implementation shifts the absolute forecast-error sequence by $H$, averages the available shifted residuals over the last $W$ samples, and uses
\begin{equation*}
r^{\mathrm{rel}}_{i,t}
=\operatorname{clip}_{[0,1]}\!\left(\exp[-40\,\bar e_{i,t}]\right),
\end{equation*}
where $\bar e_{i,t}$ is the rolling mean of these matured absolute residuals. Before any residual has matured, the global median matured error provides the frozen fallback. Thus, the reliability available at step $t$ does not use the unavailable error of the current forecast, and larger values indicate lower historical forecast error. These rules define $\mathcal{C}^{\mathrm{feas}}_{i,t}\subseteq\mathcal{C}_{i,t}$.

For each feasible candidate, A5v2 evaluates a one-step outcome with the same
trace-conditioned transition and reward equations used by the controller evaluator.
A candidate action produces an adjusted future-QoS value
$\widetilde{q}_{i,t+1}(a)$ and adjusted temporal-risk value
$\widetilde{\rho}_{i,t+1}(a)$. The corresponding model-based utility is
\begin{equation*}
\begin{aligned}
\widehat{R}_{i,t}(a)
={}&
\widetilde{q}_{i,t+1}(a)
+
0.50\!\left[
\widetilde{q}_{i,t+1}(a)-q_{i,t}
\right] \\
&-
0.10\,\widetilde{\rho}_{i,t+1}(a)
-
C_{\mathrm{res}}(a,z_{i,t})
-
C_{\mathrm{sched}}(a).
\end{aligned}
\end{equation*}
This candidate utility retains the QoS and temporal-risk structure of the controller reward in \eqref{eq:frozen_reward} and additionally subtracts the frozen action costs below. It is used for A5v2 candidate scoring and is distinct from the rollout reward recorded for controller evaluation.
The cost terms follow the frozen one-step candidate model. Table~\ref{tab:a5_action_costs}
reports the action-level values used by the implementation. The resource term contains
the recorded base cost and, for $A_1$ and $A_4$, an additional cost of $0.020$ when
$z_{i,t}=0$:
\begin{equation*}
C_{\mathrm{res}}(a,z_{i,t})
=
C_{\mathrm{res}}^{\mathrm{base}}(a)
+
0.020\,\mathbb{I}\!\left[z_{i,t}=0,\;a\in\{A_1,A_4\}\right].
\end{equation*}
Scheduler switching contributes $C_{\mathrm{sched}}(a)=0.015$ for
$A_3$, $A_4$, and $A_5$, and zero for the remaining actions.

\begin{table}[!t]
\caption{Frozen Action Costs in the A5v2 One-Step Candidate Model.}
\label{tab:a5_action_costs}
\centering
\scriptsize
\begin{tabular}{crrr}
\toprule
Action & PRB change & $C_{\mathrm{res}}^{\mathrm{base}}$ & $C_{\mathrm{sched}}$ \\
\midrule
$A_0$ & $0$     & 0.000 & 0.000 \\
$A_1$ & $+10\%$ & 0.020 & 0.000 \\
$A_2$ & $-10\%$ & 0.010 & 0.000 \\
$A_3$ & $0$     & 0.000 & 0.015 \\
$A_4$ & $+10\%$ & 0.030 & 0.015 \\
$A_5$ & $-10\%$ & 0.010 & 0.015 \\
\bottomrule
\end{tabular}
\end{table}

Here, $C_{\mathrm{res}}$ is an action-level resource-reconfiguration cost in the frozen candidate model. Its positive values for the resource-reducing actions $A_2$ and $A_5$ are therefore penalties assigned to taking those commands, rather than signed measures of net PRB consumption.

The one-step model uses the frozen risky-state bonuses and penalties reported in
Table~\ref{tab:transition_coeffs}; no transition parameter is refitted during
arbitration.

Actor preference is represented by
\begin{equation*}
P_{i,t}(a)
=
\log\!\left(
\pi_{\theta_i}(a\mid\mathbf{s}_{i,t})+10^{-12}
\right).
\end{equation*}
The forecast-reliability adjustment is
\begin{equation*}
G_{i,t}(a)
=
\begin{cases}
r^{\mathrm{rel}}_{i,t},
&
z_{i,t}=1,\; a\in\{A_1,A_3,A_4\},
\\[2pt]
-\!\left(1-r^{\mathrm{rel}}_{i,t}\right),
&
z_{i,t}=1,\; a=A_0,
\\[2pt]
-0.25,
&
z_{i,t}=0,\; a\in\{A_1,A_4\},
\\[2pt]
0,
&
\text{otherwise}.
\end{cases}
\end{equation*}
The $0.020$ safe-state term in $C_{\mathrm{res}}$ and the $-0.25$ safe-state term in $G_{i,t}$ are separate frozen mechanisms: the former belongs to one-step action cost, while the latter adjusts the candidate score using forecast reliability. Actor preference is the log-policy term defined above; the $10^{-12}$ constant is used only inside the logarithm.

The exact candidate score used by A5v2 is
\begin{equation}
\begin{array}[b]{@{}l@{}l@{}}
S_{i,t}(a)
=
\widehat{R}_{i,t}(a)
+
w_{\mathrm{actor}}P_{i,t}(a)
+
w_{\mathrm{rel}}G_{i,t}(&a),
{} \\ &
a\in\mathcal{C}^{\mathrm{feas}}_{i,t},
\end{array}
\label{eq:arbiter_score}
\end{equation}
with frozen values
\begin{equation*}
K=3,\qquad
w_{\mathrm{actor}}=0.20,\qquad
w_{\mathrm{rel}}=0.15,
\end{equation*}
and intervention margin
\begin{equation*}
\delta_{\mathrm{int}}=0.015.
\end{equation*}

A nominal score-scale analysis over 440 evaluated candidates gives mean absolute contributions of 0.1875 for $\widehat{R}_{i,t}(a)$, 0.3150 for $w_{\mathrm{actor}}P_{i,t}(a)$, and 0.0235 for $w_{\mathrm{rel}}G_{i,t}(a)$. The actor component exceeds the reward component in absolute magnitude for 85.0\% of these candidates. We therefore interpret the three terms as fixed components of the implemented arbitration score, rather than as equally scaled quantities; no weight is retuned from this analysis.

Let
\begin{equation*}
a^\star_{i,t}
=
\arg\max_{a\in\mathcal{C}^{\mathrm{feas}}_{i,t}}
S_{i,t}(a).
\end{equation*}
If the actor proposal is infeasible, the arbiter selects $a^\star_{i,t}$. When the
actor proposal is feasible, it is replaced only if
\begin{equation*}
a^\star_{i,t}\neq a^{\mathrm{actor}}_{i,t}
\quad\text{and}\quad
S_{i,t}(a^\star_{i,t})
-
S_{i,t}(a^{\mathrm{actor}}_{i,t})
\geq \delta_{\mathrm{int}}.
\end{equation*}
Otherwise, the actor proposal is retained. If all Top-$K$ candidates are infeasible,
the implementation uses $A_3$ as the fallback in a risky state and $A_0$ in a
non-risky state.

A5v2 remains a fixed execution rule around the learned actor. Algorithm~\ref{alg:a5_selection} summarizes the fixed selection path; the nominal $K=3$, actor weight 0.20, reliability weight 0.15, low-reliability cutoff 0.42, and margin 0.015 are fixed before final evaluation.

\begin{algorithm}[!t]
\caption{$M_5$ (A5v2) Reliability-Aware Top-$K$ Arbitration.}
\label{alg:a5_selection}
\begin{algorithmic}[1]
\STATE Observe $\mathbf{s}_{i,t}$ and obtain
$\pi_{\theta_i}(\cdot\mid\mathbf{s}_{i,t})$
\STATE Set $a^{\mathrm{actor}}_{i,t}\leftarrow
\arg\max_a\pi_{\theta_i}(a\mid\mathbf{s}_{i,t})$
\STATE Form $\mathcal{C}_{i,t}\leftarrow
\operatorname{TopK}(\pi_{\theta_i},K=3)$
\STATE Determine $z_{i,t}$ from
$\widehat{q}^{\,\mathrm{future}}_{i,t}$ and
$\rho^{\,\mathrm{causal}}_{i,t}$
\STATE Remove $A_2$ and $A_5$ when $z_{i,t}=1$
\STATE Remove $A_0$ when $z_{i,t}=1$ and
$r^{\mathrm{rel}}_{i,t}<0.42$
\IF{$\mathcal{C}^{\mathrm{feas}}_{i,t}=\emptyset$}
    \STATE execute $A_3$ if risky; otherwise execute $A_0$
\ELSE
    \FOR{each $a\in\mathcal{C}^{\mathrm{feas}}_{i,t}$}
        \STATE compute $\widehat{R}_{i,t}(a)$ with the frozen one-step model
        \STATE compute $P_{i,t}(a)$ and $G_{i,t}(a)$
        \STATE compute $S_{i,t}(a)$ using \eqref{eq:arbiter_score}
    \ENDFOR
    \STATE $a^\star_{i,t}\leftarrow\arg\max_a S_{i,t}(a)$
    \IF{$a^{\mathrm{actor}}_{i,t}$ is infeasible}
        \STATE execute $a^\star_{i,t}$
    \ELSIF{$a^\star_{i,t}\neq a^{\mathrm{actor}}_{i,t}$ and
    $S_{i,t}(a^\star_{i,t})-S_{i,t}(a^{\mathrm{actor}}_{i,t})\geq0.015$}
        \STATE execute $a^\star_{i,t}$
    \ELSE
        \STATE execute $a^{\mathrm{actor}}_{i,t}$
    \ENDIF
\ENDIF
\STATE Record the executed action, intervention reason, QoS, risk, recovery, and resource outcomes
\end{algorithmic}
\end{algorithm}

\subsection{Computational and Time Complexity}
\label{subsec:complexity}

The computational cost is separated into offline model construction and online decision making because forecasting-model training, PCMCI+ discovery, and policy optimization are not repeated at each control step. For the online path, $N=|\mathcal{N}|$ is the number of agents, $L$ is the forecasting look-back, $d_f$ is the input dimension at one forecast step, and $h_f$ is the LSTM hidden width. One LSTM pass over the forecasting window costs
\begin{equation*}
\mathcal{O}\!\left(Lh_f(d_f+h_f)\right),
\end{equation*}
with $L=64$ in the present controller and a fixed eight-step forecast horizon.

Each decentralized MAPPO actor uses the mapping $10\!\rightarrow\!128\!\rightarrow\!128\!\rightarrow\!6$, giving
\begin{equation*}
\mathcal{O}\!\left(10\!\cdot\!128+128^2+128\!\cdot\!6\right)
\end{equation*}
per agent and $N$ times this cost for all actors. For A5v2, $|\mathcal{A}|$ is the action-space size and $K$ is the retained Top-$K$ count. Bounded Top-$K$ selection followed by candidate filtering and scoring costs
\begin{equation*}
\mathcal{O}\!\left(|\mathcal{A}|\log K+K\right),
\end{equation*}
where $|\mathcal{A}|=6$ and $K=3$. The resulting per-decision online complexity is
\begin{equation}
\mathcal{O}\!\left(
\begin{aligned}
&Lh_f(d_f+h_f)+N(10\!\cdot\!128+128^2+128\!\cdot\!6)\\
&\quad+|\mathcal{A}|\log K+K
\end{aligned}
\right).
\end{equation}
Thus, recurrent forecasting and dense actor inference dominate the evaluated online path, while arbitration operates over a fixed six-action domain.

Offline cost depends on the number of optimization updates and conditional-independence tests. Let $B_f$ and $E_f$ denote the number of forecast-training windows processed per epoch and the number of forecast-training epochs. Forecast fitting scales as
$\mathcal{O}\!\left(E_f B_f Lh_f(d_f+h_f)\right)$ up to the corresponding backward-pass constant. For MAPPO, let $B_{\mathrm{RL}}$ be the rollout samples processed per optimization epoch and $E_{\mathrm{RL}}$ the number of optimization epochs; its optimization cost is $\mathcal{O}(E_{\mathrm{RL}}B_{\mathrm{RL}})$ times the actor--critic forward/backward cost. The reported configuration uses $B_{\mathrm{RL}}=512$ and $E_{\mathrm{RL}}=5$. For PCMCI+, let $N_{\mathrm{CI}}$ be the number of conditional-independence tests performed and $C_{\mathrm{CI}}$ the cost of one such test under its conditioning set. The discovery stage is then summarized as $\mathcal{O}(N_{\mathrm{CI}}C_{\mathrm{CI}})$. These three terms are offline and do not enter per-decision action selection.

The latency microbenchmark in Section~\ref{subsec:runtime_explainability} follows the same boundary: it measures the MAPPO actor with the A5v2 arbiter, excluding forecasting, environment execution, and rollout generation.

\section{FAROS Framework and Control Design}
\label{sec:methodology}

Forecast and temporal risk enter the state before MAPPO proposes an action. Shielding or A5v2 then governs execution without changing the common action interface.

\subsection{Reproducibility-Oriented Design Choices}
\label{subsec:repro_design}

FAROS denotes the complete forecast/risk, MAPPO, and shield/A5v2 controller. The final FAROS configuration is $M_5$, while $A_0$--$A_5$ denote the shared physical commands. The reproducibility package preserves the evaluation dataframe, configurations, matched starts, environment source, stored forecast path, and ten hash-verified validation-selected checkpoints. A clean ten-seed evaluation reproduces the reported $M_5$ metrics. Because upstream clustering and the original forecast-training lineage are incomplete, Section~\ref{subsec:forecast_results} validates forecasting with a separate leakage-safe LSTM without replacing the stored controller forecast.

\section{Experimental Methodology}

\label{sec:experimental_methodology}

The evaluation follows the FAROS decision path under protected temporal splitting. Trace-conditioned controller results and independent ns-3 action-level evidence remain separate.

Forecast and later-QoS vulnerability tests precede controller comparisons. Variants share paired seeds, matched start indices, common actions, and frozen transition assumptions; sensitivity and external validation are separate.

\subsection{Dataset Construction and Temporal Evaluation Setup}
\label{subsec:dataset}

The controller data come from the public Open RAN Commercial Traffic Twinning Dataset, which contains commercial LTE traffic twinned on an experimental Open RAN platform. The underlying Open RAN experimentation and data-collection setting follows the OpenRAN Gym platform described by Bonati \emph{et al.}~\cite{bonati2023openran_gym}. FAROS uses the retained \texttt{slicing\_1} dataset with three cluster labels that define the MAPPO agents. Scheduler codes 0 and 2 are preserved; because the retained dataset lacks an authoritative code-to-name mapping, numerical claims do not assign them to round-robin or PF. $A_3$ applies the stored scheduler-switch operation. Table~\ref{tab:dataset_summary} summarizes the fixed dataset and evaluation dimensions: the aligned dataframe contains 1,001,456 observations and 50 columns across 184 \texttt{(cluster,reservation)} series, with cluster populations 246,674, 421,017, and 333,765. Upstream cluster construction is not preserved, so reproducibility begins from these retained assignments. The traces represent experimentally twinned demand rather than a live O-RAN deployment.

\begin{table}[!t]
\caption{Controller Dataset and Evaluation Dimensions.}
\label{tab:dataset_summary}
\centering
\small
\begin{tabular}{lr}
\toprule
Quantity & Fixed value \\
\midrule
Controller observations & 1,001,456 \\
Controller columns & 50 \\
Number of agents & 3 \\
Unique cluster--reservation series & 184 \\
Scheduling configurations & 2 \\
\texttt{cluster\_1} observations & 246,674 \\
\texttt{cluster\_2} observations & 421,017 \\
\texttt{cluster\_3} observations & 333,765 \\
Training observations & 518,430 \\
State dimension & 10 \\
Action dimension & 6 \\
Forecast look-back & 64 steps \\
Forecast horizon & 8 steps \\
Evaluation seeds & 10 (0--9) \\
Matched start indices & 500 \\
Matched-risk horizon & 10 steps \\
\bottomrule
\end{tabular}
\end{table}

\begin{figure*}[!t]
\centering
\begin{tikzpicture}[
    x=1mm,
    y=1mm,
    >=Latex,
    font=\footnotesize,
    info/.style={
        draw=black!85,
        fill=white,
        rounded corners=1.2pt,
        line width=0.72pt,
        align=center,
        text width=36mm,
        minimum height=9.5mm,
        inner sep=1.5pt
    },
    cluster/.style={
        draw=black!80,
        fill=white,
        rounded corners=1.2pt,
        line width=0.68pt,
        align=center,
        text width=28mm,
        minimum height=9mm,
        inner sep=1.4pt
    },
    phase/.style={
        draw=black!90,
        fill=white,
        rounded corners=1.2pt,
        line width=0.78pt,
        align=center,
        text width=57mm,
        minimum height=11.5mm,
        inner sep=1.6pt
    },
    contract/.style={
        draw=black!80,
        fill=white,
        rounded corners=1.2pt,
        line width=0.68pt,
        align=center,
        text width=151mm,
        minimum height=7.5mm,
        inner sep=1.2pt
    },
    arrow/.style={
        -{Latex[length=1.7mm,width=1.1mm]},
        draw=black!85,
        line width=0.70pt,
        shorten >=1.2pt,
        shorten <=1.2pt
    },
    thin/.style={
        draw=black!65,
        line width=0.60pt
    },
    gap/.style={
        draw=black!70,
        densely dashed,
        line width=0.66pt
    }
]

\node[info] (dataset) at (-63,15) {
    \textbf{Controller Evaluation Dataset}\\
    1,001,456 observations, 50 columns
};

\node[cluster] (c1) at (-19,15) {
    \textbf{\texttt{cluster\_1}}\\
    246,674 observations
};

\node[cluster] (c2) at (20,15) {
    \textbf{\texttt{cluster\_2}}\\
    421,017 observations
};

\node[cluster] (c3) at (59,15) {
    \textbf{\texttt{cluster\_3}}\\
    333,765 observations
};

\draw[arrow] (dataset.east) -- (c1.west);
\draw[thin] (c1.east) -- (c2.west);
\draw[thin] (c2.east) -- (c3.west);

\node[phase] (val) at (-43,-1) {
    \textbf{Validation}\\
    128 starts, indices 172,810--202,239
};

\node[phase] (test) at (43,-1) {
    \textbf{Protected Final Test}\\
    500 starts, indices 202,290--246,627
};

\draw[gap]
    ($(val.east)+(1.8,0)$) --
    ($(test.west)+(-1.8,0)$);

\node[
    fill=white,
    inner xsep=1.8pt,
    inner ysep=0.5pt,
    font=\footnotesize\bfseries,
    text=black!75,
    align=center
] at (0,-1) {
    Protected start-index separation\\
    Start overlap $=0$
};

\node[contract] (contract) at (0,-15) {
    \textbf{Frozen paired-evaluation contract:}
    seeds 0--9; 10-step matched-risk horizon; no final-test selection; same six-action interface, protected starts, and frozen controller conditions
};

\draw[arrow] ($(val.south)+(12,0)$) -- ($(contract.north)+(-28,0)$);
\draw[arrow] ($(test.south)+(-12,0)$) -- ($(contract.north)+(28,0)$);

\end{tikzpicture}

\caption{Dataset construction and protected temporal evaluation protocol used for the FAROS controller evaluation.}
\label{fig:data_pipeline}
\end{figure*}
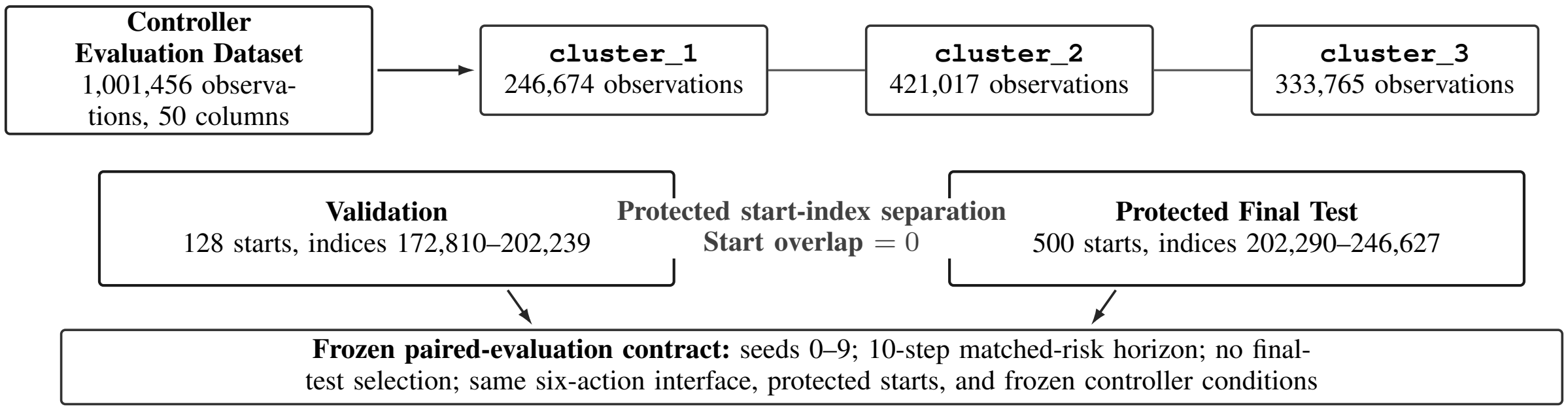

Fig.~\ref{fig:data_pipeline} summarizes the protected temporal split and paired-evaluation protocol. The 64-step context, eight-step forecast, and 10-step rollout are reported in decision steps because no physical sampling period is retained; temporal order and reservation boundaries are preserved. Training uses 172,810 observations per agent (518,430 total); validation uses 128 starts at indices 172810--202239 and final evaluation 500 starts at 202290--246627, with no start overlap. Historical context may precede an evaluation start, but validation/test targets and post-start outcomes are excluded from training and checkpoint selection. Seven reset-eligible positions (172801--172807) remain inside training.

\subsection{Matched-Risk Evaluation Contract}
\label{subsec:evaluation_contract}

The frozen evaluation uses seeds 0--9 and the same 500 unique start indices per seed selected under the risky-start protocol, fixed once with matched-start seed 2026; the retained start file and evaluation snapshot are byte-identical. With three agents and a 10-step horizon, each learned-controller evaluation has 15,000 agent trajectories and 150,000 decisions. These conditioned rates do not represent full-trace prevalence. Starts, actions, and transition assumptions remain fixed. Thinning to 490 non-overlapping 10-step starts preserves all eight paired $M_5$--$M_4$ effect directions. Table~\ref{tab:evaluation_contract} records the frozen A5v2 parameters; displayed rounding does not alter machine-precision values, and no threshold uses the final test.

\begin{table}[!t]
\caption{Matched-Risk Evaluation Contract.}
\label{tab:evaluation_contract}
\centering
\small
\begin{tabular}{lr}
\toprule
Property & Value \\
\midrule
Seeds & 0--9 \\
Matched start indices & 500 \\
Evaluation horizon & 10 steps \\
Agent-level trajectories & 15,000 \\
Recorded decisions & 150,000 \\
$M_5$/A5v2 low-QoS threshold & 0.064818 \\
$M_5$/A5v2 high-risk threshold & 0.442658 \\
A5v2 Top-K & 3 \\
Actor weight & 0.20 \\
Reliability weight & 0.15 \\
Low-reliability cutoff & 0.42 \\
Intervention margin & 0.015 \\
\bottomrule
\end{tabular}
\end{table}

\subsection{Fixed Model-Matched Deterministic Reference}
\label{subsec:deterministic_rule}

The fixed deterministic reference tests direct threshold logic under the same information and evaluator as $M_5$. It uses the six-action interface in Table~\ref{tab:action_space}, with thresholds
\begin{equation*}
\tau_q^{(\mathrm{eval})}\approx0.064818,
\qquad
\tau_\rho^{(\mathrm{eval})}\approx0.442658.
\end{equation*}
These displayed boundaries are rounded to six decimal places; evaluation retains the same machine-precision thresholds as final $M_5$/A5v2, fixed without final-test tuning. The rule has no learned parameters or random seed, so it is evaluated once without an across-seed confidence interval.

With current QoS $q_t$, predicted future QoS $\widehat{q}^{\,\mathrm{future}}_t$, and temporal risk $\rho^{\,\mathrm{causal}}_t$, the rule evaluates
\begin{align*}
b_q &= \mathbb{I}\!\left[\widehat{q}^{\,\mathrm{future}}_t<\tau_q^{(\mathrm{eval})}\right], \\
b_\rho &= \mathbb{I}\!\left[\rho^{\,\mathrm{causal}}_t>\tau_\rho^{(\mathrm{eval})}\right], \\
b_f &= \mathbb{I}\!\left[\widehat{q}^{\,\mathrm{future}}_t<q_t\right], \\
b_d &= \mathbb{I}\!\left[q_t-\widehat{q}^{\,\mathrm{future}}_t\geq0.02\right].
\end{align*}
Algorithm~\ref{alg:deterministic_rule} applies these conditions in fixed priority and therefore uses only $A_0$, $A_1$, and $A_4$ from the common six-action space. It never issues the resource-reduction commands $A_2$ or $A_5$; any decrease-like quantity from another logging view must therefore not be interpreted as a rule-selected reduction action.

\begin{algorithm}[!t]
\caption{Fixed Model-Matched Deterministic Reference.}
\label{alg:deterministic_rule}
\begin{algorithmic}[1]
\STATE Observe $q_t$, $\widehat{q}^{\,\mathrm{future}}_t$, and $\rho^{\,\mathrm{causal}}_t$
\STATE $b_q\leftarrow(\widehat{q}^{\,\mathrm{future}}_t<\tau_q)$
\STATE $b_\rho\leftarrow(\rho^{\,\mathrm{causal}}_t>\tau_\rho)$
\STATE $b_f\leftarrow(\widehat{q}^{\,\mathrm{future}}_t<q_t)$
\STATE $b_d\leftarrow(q_t-\widehat{q}^{\,\mathrm{future}}_t\geq0.02)$
\IF{$b_\rho\land(b_q\lor b_d)$}
    \STATE execute $A_4$
\ELSIF{$b_q\lor b_\rho\lor b_f$}
    \STATE execute $A_1$
\ELSE
    \STATE execute $A_0$
\ENDIF
\end{algorithmic}
\end{algorithm}

The rule is a model-matched diagnostic rather than an external or reproduced published baseline.

\subsection{Independent ns-3 Validation Protocol}
\label{subsec:ns3_methodology}

The ns-3 study checks the six-action interface rather than replaying FAROS. Twenty matched runs execute $A_0$--$A_5$ with \texttt{RngSeed}=2026 and matched \texttt{RngRun}, giving 120 successful executions with all ten controller seeds represented in provenance. Balanced action sampling is separate from the natural $M_5$ distribution, and simulator outcomes do not affect training, selection, thresholds, or arbitration. Goodput, PDR, packet loss, and latency test packet-level command effects; the experiment does not reproduce sequential state-conditioned $M_5$ behavior.

\subsection{Baselines and Metrics}
\label{subsec:baselines_metrics}

Forecast references are persistence and a past-only eight-sample moving average (MovingAverage8). They are controlled references linked to the forecasting literature in Section~\ref{sec:related_work}, not reproductions of those studies \cite{phyu2023multislice_forecasting,perifanis2023federated_forecast,kasuluru2025probabilistic_oran}. The independent validation uses seed 42, a one-layer 64-unit LSTM, batch size 1024, learning rate $10^{-3}$, at most 30 epochs, and patience 5. The checkpoint is selected only by validation MAE; the test partition is not used for model selection.

The controller comparison fixes 500 protected starts per seed, horizon, actions, and transition--reward equations. $M_4$ executes the Forecast+Temporal-Risk actor directly. $M_4$+Shield applies only the frozen $M_5$/A5v2 risky-state $A_2/A_5$ prohibition to the same actors; it does not use the $A_0$ reliability rule, Top-$K$ scoring, or margin-gated arbitration. $M_5$ applies A5v2 to the same-seed actor. Fair MAPPO masks forecast and risk, is trained independently for 30 iterations rather than the FAROS lineage's 60, and is therefore a common-dynamics reference rather than a training-budget-matched ablation. Learned controllers use ten seeds with training seed as the inference unit. The deterministic rule is one fixed point, contextualized by proactive and constrained O-RAN control \cite{staffolani2024prorl,nagib2024safe_accelerated_oran_slicing} without claiming reproduction.

Forecast accuracy uses MAE, RMSE, $R^2$, and Pearson correlation. Controller statistics are first computed per training seed and then summarized across the ten seeds. For seed $s$, let $\mathcal{I}_{\mathrm{risk}}^{(s)}=\{n:z_n=1\}$ and $N_{\mathrm{risk}}^{(s)}=|\mathcal{I}_{\mathrm{risk}}^{(s)}|$. The model-internal action-relative changes are $\Delta q_n=\widehat{q}^{\,\mathrm{future,adj}}_n-\widehat{q}^{\,\mathrm{future}}_n$ and $\Delta\rho_n=\rho_n^{\mathrm{post}}-\rho_n^{\mathrm{pre}}$. The reported risky-state QoS change is
\begin{equation*}
\overline{\Delta q}_{\mathrm{risk}}^{(s)}=\frac{1}{N_{\mathrm{risk}}^{(s)}}\sum_{n\in\mathcal{I}_{\mathrm{risk}}^{(s)}}\Delta q_n,
\end{equation*}
with the analogous risky-state temporal-risk change. For binary indicators $p_n$, $h_n$, and $v_n$ denoting proactive execution, prohibited execution, and intervention,
\begin{equation*}
r_{\mathrm{pro|risk}}^{(s)}=\frac{\sum_{n\in\mathcal{I}_{\mathrm{risk}}^{(s)}}p_n}{N_{\mathrm{risk}}^{(s)}},\quad
r_{\mathrm{proh|risk}}^{(s)}=\frac{\sum_{n\in\mathcal{I}_{\mathrm{risk}}^{(s)}}h_n}{N_{\mathrm{risk}}^{(s)}}.
\end{equation*}
``Proactive'' means $A_1$, $A_3$, or $A_4$. Intervention is reported over all decisions and, where stated, conditional on risky states. Recovery is computed per seed over initially risky agent trajectories as the fraction recovered within the fixed ten-step horizon; recovery steps average only recovered trajectories. Reward and realized resource changes use post-transition records, so clipped commands are not credited as realized PRB changes. Proposal, execution, and outcome metrics remain distinct.

Across-seed summaries report the sample mean, sample standard deviation, and two-sided 95\% confidence interval. Paired $M_5$--$M_4$ inference uses the training seed as the resampling unit and reports the mean difference, a 95\% percentile bootstrap interval over the ten seed pairs, and paired Cohen's $d_z$. Paired $t$ and Wilcoxon tests are secondary, with Holm correction across eight controller metrics. For ten nonzero pairs, the minimum two-sided exact Wilcoxon value is $2/2^{10}=0.001953$; interpretation therefore rests on effect estimates and uncertainty intervals. Counts and proposal-to-execution mismatches are checked over the complete 150,000-decision trace.

\section{Results}

\label{sec:results}

Results use the frozen protocol of Section~\ref{sec:experimental_methodology}: ten checkpoints, 500 matched start indices per seed, a 10-step horizon, and 150,000 decisions per learned controller. The fixed diagnostic has 15,000 decisions. SHA-256 manifests fix the evaluated files; forecast, runtime, and ns-3 tests use 138,212 sequences, 3,000 post-warm-up decisions per method, and 20 matched runs per action, respectively.

\subsection{Forecast Accuracy}
\label{subsec:forecast_results}

\begin{figure}[!t]
    \centering
    \includegraphics[width=\columnwidth]{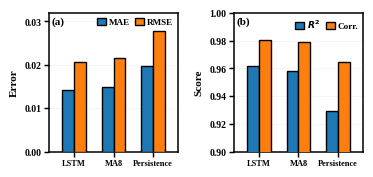}
    \caption{Independent leakage-safe $H=8$ forecast comparison on 138,212 protected test sequences. The LSTM has the lowest MAE/RMSE and highest $R^2$/correlation relative to MovingAverage8 and persistence; exact values are shown in the plot.}
    \label{fig:forecast_metrics}
\end{figure}

Table~\ref{tab:forecast_baselines} and Fig.~\ref{fig:forecast_metrics} make the $H=8$ baseline comparison explicit on 138,212 protected test sequences. The validation-selected LSTM has lower MAE than MovingAverage8 (0.014279 vs. 0.014994) and persistence (0.019794), while reaching the highest $R^2$ (0.961723). Across 156 reservation groups, its MAE advantage over MovingAverage8 is 0.000683 with 95\% CI [0.000597, 0.000766], and 148 groups favor the LSTM. This supports $H=8$ forecasting on the retained traces; it does not validate the historical model that produced the stored FAROS forecast.

\begin{table*}[!t]
\caption{Independent Leakage-Safe Forecast Validation.}
\label{tab:forecast_baselines}
\centering
\small
\begin{tabular}{lccccc}
\toprule
Method & $n$ & MAE & RMSE & $R^2$ & Corr. \\
\midrule
Independent LSTM & 138,212 & \textbf{0.014279} & \textbf{0.020560} & \textbf{0.961723} & \textbf{0.980721} \\
MovingAverage8 (past-only) & 138,212 & 0.014994 & 0.021605 & 0.957732 & 0.978671 \\
Persistence & 138,212 & 0.019794 & 0.027913 & 0.929450 & 0.964750 \\
\bottomrule
\end{tabular}
\end{table*}

A paired evaluation-time perturbation compares the fixed nominal package only with a verified shuffled-forecast condition. Shuffling raises mean reward from 0.281455 to 0.319659, while model-internal action-relative QoS change falls from 0.097039 to 0.084670, risk change moves from $-0.146634$ to $-0.129792$, and recovery falls from 0.953009 to 0.925131. Because forecast-dependent terms enter \eqref{eq:frozen_reward}, the model-based reward is not a standalone operational metric; we interpret it with QoS change, risk change, and recovery. No forecast-disabled $M_4$ checkpoint is available, so this is an evaluation-time sensitivity test rather than a trained ablation.

\subsection{Incremental Vulnerability Information from Temporal Risk}
\label{subsec:causal_results}

\begin{figure}[!t]
    \centering
    \includegraphics[width=\columnwidth]{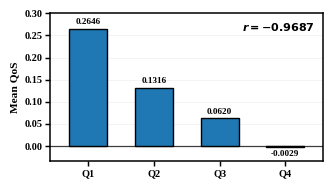}
    \caption{Mean QoS across numerical quartiles of the temporal-risk score on the 1,001,456 aligned controller observations. Q1 contains the lowest score values and Q4 the highest; Pearson correlation is $r=-0.9687$.}
    \label{fig:risk_quartiles}
\end{figure}

On the aligned controller population, mean QoS decreases monotonically across the numerical temporal-risk quartiles, from 0.264621 in Q1 to $-0.002944$ in Q4, with Pearson correlation $r=-0.9687$ (Fig.~\ref{fig:risk_quartiles}). This correlation is descriptive and is not treated as independent evidence for the risk construction. The retained risk score uses granted PRBs and downlink MCS, while the trace-derived QoS utility itself contains granted PRBs. Part of the association can therefore arise mechanically from shared state variables. The subsequent future-vulnerability test uses later observed QoS and is reported separately.

\begin{figure}[!t]
    \centering
    \includegraphics[width=\columnwidth]{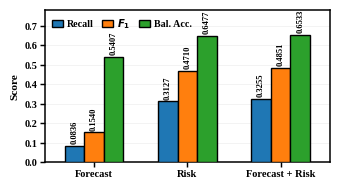}
    \caption{Future-QoS vulnerability comparison at $H=3$. Forecast+temporal risk improves recall (0.083592 to 0.325462) and balanced accuracy (0.540750 to 0.653257) over forecast only, with FPR 0.018947.}
    \label{fig:causal_detector_plot}
\end{figure}

Table~\ref{tab:causal_detector} and Fig.~\ref{fig:causal_detector_plot} compare future-vulnerability detectors using subsequently observed QoS. A positive target crosses $\tau_q^{(M_5)}$ within the prespecified $H=3$ horizon. Among 149,980 targets, prevalence is 53.761\%; the always-positive F1 of 0.699280 is non-discriminative because balanced accuracy is 0.5. Relative to forecast only, adding temporal risk raises recall from 0.083592 to 0.325462 and balanced accuracy from 0.540750 to 0.653257, with FPR 0.018947. We therefore emphasize balanced accuracy and error rates rather than F1.

\begin{table*}[!t]
\caption{Future-QoS Vulnerability Diagnostic at $H=3$.}
\label{tab:causal_detector}
\centering
\small
\begingroup
\setlength{\tabcolsep}{4pt}
\begin{tabular}{@{}lrrrrrr@{}}
\toprule
Detector & Precision & Recall & F1 & Balanced acc. & FPR & FNR \\
\midrule
Always positive & 0.537610 & 1.000000 & 0.699280 & 0.500000 & 1.000000 & 0.000000 \\
Forecast only & 0.978940 & 0.083592 & 0.154031 & 0.540750 & 0.002091 & 0.916408 \\
Temporal risk only & 0.954599 & 0.312663 & 0.471043 & 0.647687 & 0.017289 & 0.687337 \\
Forecast + temporal risk & 0.952315 & 0.325462 & 0.485127 & 0.653257 & 0.018947 & 0.674538 \\
\bottomrule
\end{tabular}
\endgroup
\end{table*}

The gain supports short-horizon vulnerability discrimination rather than intervention causality. Of 20,671 forecast-only false negatives flagged by temporal risk, 94.3447\% later become vulnerable; this remains observational.

\subsection{Matched-Risk Controller Comparison}
\label{subsec:unified_lineage_ablation}

Table~\ref{tab:unified_lineage_ablation} compares Fair MAPPO, $M_4$, shield-controlled $M_4$, $M_5$, and the fixed model-matched diagnostic under the common matched-risk contract. Fair MAPPO shows substantial seed variability and remains a common-dynamics reference rather than a budget-matched ablation; its model-internal QoS-change quantity is not interpreted as measured network improvement.

Applying only the frozen risky-state $A_2/A_5$ prohibition to $M_4$ removes its 6.67\% prohibited-action rate. Under the same evaluator, $M_5$ has mean reward 0.277807 versus 0.246486 for $M_4$+Shield and ten-step recovery 90.56\% versus 80.87\%. Across ten matched seeds, the paired differences are 0.031322 for reward (95\% CI [0.008173, 0.054470]) and 0.096927 for recovery ([0.043174, 0.150680]); recovery steps are essentially unchanged. The shield intervenes on 33,347 of 150,000 $M_4$ decisions (22.23\%), separating enforced admissibility from the additional outcome differences associated with A5v2 arbitration.

\begin{table*}[!t]
\caption{Matched-risk controller comparison. Learned controllers report ten-seed mean $\pm$ SD; the model-matched rule is one fixed point. Fair MAPPO is the independently trained common-dynamics reference with forecast/risk masked and no shield.}
\label{tab:unified_lineage_ablation}
\centering
\scriptsize
\setlength{\tabcolsep}{2.3pt}
\resizebox{\textwidth}{!}{%
\begin{tabular}{lccccc}
\toprule
Metric & Fixed rule & Fair MAPPO & $M_4$ & $M_4$+Shield & $M_5$: FAROS \\
\midrule
Mean local reward
& 0.301505
& $0.037007\pm0.168395$
& $0.188753\pm0.076943$
& $0.246486\pm0.040369$
& $0.277807\pm0.012672$ \\

Prohibited-action rate, risky states
& 0.000000
& $0.242596\pm0.196296$
& $0.066729\pm0.048296$
& $0.000000\pm0.000000$
& $0.000000\pm0.000000$ \\

Proactive rate, risky states
& 1.000000
& $0.660099\pm0.199291$
& $0.822766\pm0.144611$
& $0.877987\pm0.123849$
& $0.982330\pm0.022899$ \\

Recovery within ten steps
& 0.975672
& $0.547119\pm0.151451$
& $0.655058\pm0.135045$
& $0.808707\pm0.071807$
& $0.905634\pm0.047492$ \\

Recovery steps, recovered only
& 3.299213
& $2.908149\pm0.411096$
& $2.850026\pm0.171742$
& $3.205451\pm0.273738$
& $3.200024\pm0.076056$ \\
\bottomrule
\end{tabular}%
}
\end{table*}

The fixed model-matched diagnostic reaches reward 0.301505 and recovery 97.57\%, above FAROS; its evaluator-matched thresholds and action-response equations preclude a FAROS-superiority claim over direct threshold logic. The shield control confirms that zero prohibited actions arise from predicate enforcement.

A same-state diagnostic compares A5v2 with the model-matched diagnostic rule on states encountered along the frozen A5v2 trajectories. Their executed actions differ in 77.73\% of paired decisions. This is a decision-level comparison on a common state rather than a trajectory-level disagreement rate, because later states can diverge once different actions are executed. We use the value only to show that A5v2 does not mechanically reproduce the rule, without interpreting disagreement as evidence of better performance.

\subsection{Actor-Proposal and Execution Audit}
\label{subsec:rule_results}

\begin{figure}[!t]
    \centering
    \includegraphics[width=\columnwidth]{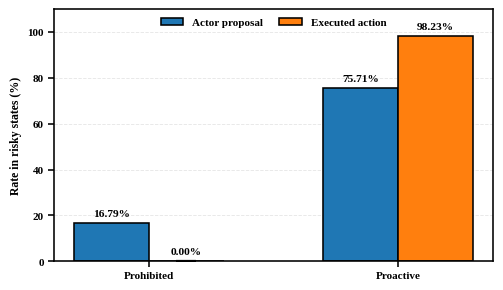}
    \caption{Actor proposals and executed actions in risky states. A5v2 execution filtering reduces the prohibited-action rate from 16.79\% to 0\%, while the proactive-action rate increases from 75.71\% to 98.23\%.}
    \label{fig:rule_behavior}
\end{figure}

Table~\ref{tab:resource_replay} and Fig.~\ref{fig:rule_behavior} audit each A5v2 actor argmax against its final command, separately from the $M_4$ rollout. Here, ``selected'' denotes the post-arbitration command, distinct from the actor proposal. The arbiter changes 22.61\% of proposals. In risky states, A5v2 execution filtering reduces the prohibited-action rate from 16.79\% at proposal time to 0\% at execution, while the proactive-action rate rises from 75.71\% to 98.23\%.

\begin{table*}[!t]
\caption{Actor-proposal and final-execution audit over the frozen $M_5$ trace. The trace contains 30,114 risky decisions in total; prohibited/proactive rate entries are unweighted means of the ten seed-level risky-state rates, consistent with Table~\ref{tab:baseline_statistics}, rather than pooled 30,114-decision proportions. The zero executed prohibited rate is predicate-enforced.}
\label{tab:resource_replay}
\centering
\small
\setlength{\tabcolsep}{6pt}
\renewcommand{\arraystretch}{1.08}
\begin{tabular*}{\textwidth}{@{\extracolsep{\fill}}lrl@{}}
\toprule
Metric & Value & Audit interpretation \\
\midrule
Actor prohibited rate, risky states & 16.7905\% & Proposal \\
Executed prohibited rate, risky states & 0.0000\% & Predicate-enforced execution \\
Actor proactive rate, risky states & 75.7066\% & Proposal \\
Executed proactive rate, risky states & 98.2330\% & Execution \\
Proposal replacement rate, all decisions & 22.6073\% & 33,911 decisions \\
Intervention rate, risky states & 31.3280\% & Arbiter intervention \\
Selected-to-executed mismatch & 0.0000\% & 0 decisions \\
\bottomrule
\end{tabular*}
\end{table*}

Across all 150,000 executed $M_5$ decisions, 40.27\% use a resource-increasing command, 23.26\% use a resource-decreasing command, and 36.46\% are resource-neutral. Under the action semantics of Table~\ref{tab:action_space}, these frequencies correspond to an expected multiplicative resource factor of 1.0170, or an action-implied net change of +1.70\%. This is a command-level resource proxy. The evaluator does not record physical PRB utilization or over-allocation, so we do not interpret it as measured resource efficiency.

\subsection{Across-Seed Execution Stability}
\label{subsec:paired_statistics}

Table~\ref{tab:baseline_statistics} reports proposal-to-execution variability. A5v2 execution filtering blocks prohibited proposals for every seed, executed proactive actions exceed 94.20\%, and no post-arbitration selected command differs from execution.

\begin{table*}[!t]
\caption{Across-seed statistics for the actor-to-execution audit.}
\label{tab:baseline_statistics}
\centering
\scriptsize
\begin{tabular}{lrrrrr}
\toprule
Metric & Mean & SD & 95\% CI & Minimum & Maximum \\
\midrule
Actor prohibited rate, risky states & 0.167905 & 0.123938 & [0.079245, 0.256565] & 0.002766 & 0.355324 \\
Executed prohibited rate, risky states & 0.000000 & 0.000000 & [0.000000, 0.000000] & 0.000000 & 0.000000 \\
Actor proactive rate, risky states & 0.757066 & 0.164647 & [0.639285, 0.874847] & 0.484811 & 0.961964 \\
Executed proactive rate, risky states & 0.982330 & 0.022899 & [0.965949, 0.998710] & 0.942007 & 1.000000 \\
Intervention rate, all decisions & 0.226073 & 0.077578 & [0.170578, 0.281569] & 0.130467 & 0.377533 \\
Intervention rate, risky states & 0.313280 & 0.134644 & [0.216961, 0.409599] & 0.104159 & 0.518871 \\
\bottomrule
\end{tabular}
\end{table*}

\begin{figure*}[!t]
    \centering
    \includegraphics[width=0.68\textwidth]{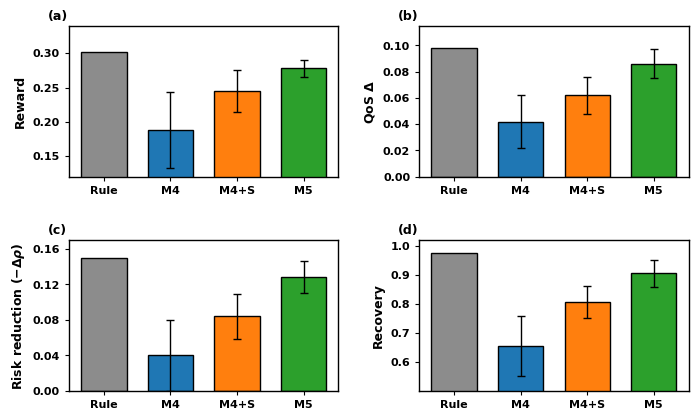}
    \caption{Matched-risk controller comparison under the common frozen contract: (a) reward, (b) model-internal risky-state QoS change, (c) risky-state temporal-risk reduction $-\Delta\rho$, and (d) ten-step recovery. $M_5$ improves reward and recovery over $M_4$ and $M_4$+Shield, while the fixed model-matched rule remains higher. Error bars are 95\% CIs for learned controllers; the rule is a point estimate.}
    \label{fig:baseline_statistics}
\end{figure*}

Fig.~\ref{fig:baseline_statistics} makes the matched-risk comparison visible across four outcomes. Relative to $M_4$+Shield, $M_5$ increases reward from 0.24649 to 0.27781 and recovery from 80.87\% to 90.56\%; the fixed model-matched rule remains higher at 0.30151 and 97.57\%.

\subsection{Local Arbitration-Parameter Sensitivity}
\label{subsec:transition_sensitivity}

Table~\ref{tab:transition_sensitivity} tests one-at-a-time sensitivity around the frozen A5v2 setting without retraining; checkpoints, starts, transition--reward model, thresholds, and other parameters remain fixed. Across $w_{\mathrm{actor}}\in\{0.15,0.25\}$ and $\delta_{\mathrm{int}}\in\{0.010,0.020\}$, each evaluated over ten seeds, prohibited actions remain zero, reward ranges from 0.275543 to 0.280119, and recovery from 0.896927 to 0.910755. These results support local stability around the nominal setting, not global parameter insensitivity.

\begin{table*}[!t]
\caption{Local sensitivity of A5v2 arbitration parameters around the frozen nominal configuration. The nominal row is the protected $M_5$ evaluation; each perturbation varies one parameter at a time over ten seeds without retraining.}
\label{tab:transition_sensitivity}
\centering
\scriptsize
\setlength{\tabcolsep}{3.2pt}
\begin{tabular}{lccccccc}
\toprule
Setting & Reward & Prohibited, risky & Proactive, risky & Model QoS $\Delta$, risky & Risk $\Delta$, risky & Recovery & Recovery steps \\
\midrule
Nominal: $w_{\mathrm{actor}}=0.20,\ \delta_{\mathrm{int}}=0.015$
& 0.277807 & 0.000000 & 0.982330 & 0.086186 & -0.128903 & 0.905634 & 3.200024 \\
$w_{\mathrm{actor}}=0.15$
& 0.280119 & 0.000000 & 0.982282 & 0.087264 & -0.130527 & 0.910755 & 3.206772 \\
$w_{\mathrm{actor}}=0.25$
& 0.275543 & 0.000000 & 0.982116 & 0.084707 & -0.126622 & 0.896927 & 3.178952 \\
$\delta_{\mathrm{int}}=0.010$
& 0.278371 & 0.000000 & 0.982329 & 0.086304 & -0.129080 & 0.906274 & 3.201211 \\
$\delta_{\mathrm{int}}=0.020$
& 0.277172 & 0.000000 & 0.982329 & 0.086001 & -0.128620 & 0.904994 & 3.200086 \\
\bottomrule
\end{tabular}
\end{table*}

\subsection{Evaluation Integrity and Reproducibility}
\label{subsec:threshold_sensitivity}

Table~\ref{tab:margin_sensitivity} records the main integrity checks. Dataset, starts, configuration, source, and seed-specific checkpoint identities are fixed before evaluation. A clean run over all ten frozen checkpoints, without training or parameter selection, reproduces the reported $M_5$ metrics. Seven reset-eligible positions may truncate training episodes, but all remain inside the training region and cannot expose validation or final-test observations.

\begin{table}[!t]
\caption{Integrity Checks for the Fixed Evaluation Protocol.}
\label{tab:margin_sensitivity}
\centering
\scriptsize
\begin{tabular}{lr}
\toprule
Check & Verified value \\
\midrule
Training in evaluation run & No \\
Seed checkpoints & 10 \\
Matched starts & 500 unique \\
Expected decisions per seed & 15,000 \\
Total recorded decisions & 150,000 \\
Total agent-level trajectories & 15,000 \\
Selected/executed mismatches & 0 \\
\bottomrule
\end{tabular}
\end{table}

\subsection{Evidence Boundary}
\label{subsec:ood_results}

Table~\ref{tab:load_stress} bounds the evidence for the retained trace population under the fixed transition contract. A separate trace-conditioned offered-load sensitivity test scales the retained load representation by $0.25\times$, $0.50\times$, $0.75\times$, $1.00\times$, and $1.25\times$, using ten seeds, 256 matched starts, and a ten-step horizon. The observed stability supports only this tested load range within the trace-conditioned model. It does not establish robustness to larger distribution shifts, forecast corruption, severe temporal-risk shift, or growth in the number of agents, slices, or actions.

\begin{table*}[!t]
\caption{Scope of the Reported Evidence.}
\label{tab:load_stress}
\centering
\small
\setlength{\tabcolsep}{4.5pt}
\renewcommand{\arraystretch}{1.10}
\begin{tabular}{p{3.2cm}p{5.4cm}p{7.4cm}}
\toprule
Evidence source & Supported statement & Boundary \\
\midrule
Independent leakage-safe forecast test & $H=8$ forecast accuracy on protected test sequences & Validation model is separate from the stored FAROS forecast path \\
Matched-risk A5v2 trace & Ten-seed outcome and execution audit & Trace-conditioned transition and reward model \\
Offered-load sensitivity & Stable behavior over $0.25\times$--$1.25\times$ tested load & Trace-conditioned load robustness, not computational scaling \\
Symbolic predicate audit & No prohibited final actions in risky states & Predicate compliance, not formal safety \\
Balanced matched ns-3 action runs & Packet-level direction of six commands & Action-level evidence, not sequential FAROS validation \\
\bottomrule
\end{tabular}
\end{table*}

\subsection{Runtime, Reliability, and Explainability}
\label{subsec:runtime_explainability}

\begin{figure}[!t]
    \centering
    \includegraphics[width=\columnwidth]{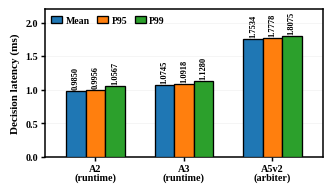}
    \caption{Controller-side actor/arbitration runtime for Runtime-Cfg2, Runtime-Cfg3, and A5v2. Runtime-Cfg2 and Runtime-Cfg3 are configuration names, unrelated to action symbols $A_2$ and $A_3$. Bars report mean, P95, and P99 latency over 3,000 decisions after 200 warm-up decisions.}
    \label{fig:deployment_evidence}
\end{figure}

Fig.~\ref{fig:deployment_evidence} reports controller-side actor/arbitration runtime for two retained pre-A5v2 configurations, denoted Runtime-Cfg2 and Runtime-Cfg3, and for A5v2. After 200 warm-up decisions, 3,000 decisions per method give mean/P95/P99 latency of 0.985050/0.995561/1.056689 ms, 1.074527/1.091840/1.128021 ms, and 1.753396/1.777802/1.807549 ms, respectively. The benchmark excludes E2 transport, telemetry, RIC scheduling, and actuator delay and is not end-to-end O-RAN latency.

The 30-row reliability audit relates $r^{\mathrm{rel}}$ to forecast error across six conditions and five bins. In 1,500 balanced actor explanations, mean absolute SHAP attribution is highest for temporal risk (0.186897), predicted future QoS (0.146218), downlink MCS (0.097787), and granted PRBs (0.075653), following actor-level explainable-RL practice \cite{xu2024xrl_shap_cache}. Final-command interpretation uses a separate 4,500-row candidate audit with 263 interventions. Parameter randomization retains rank correlations 0.804878, 0.721212, and 0.830303 across clusters, so SHAP is treated only as descriptive actor attribution.

\section{External ns-3 Action-Level Validation}
\label{sec:ns3_validation}

The independent ns-3 study tests the six-action interface outside the trace-conditioned evaluator. Twenty matched \texttt{RngRun} values execute all commands, giving 120 successful paired executions with all ten controller seeds represented in provenance. Simulator outcomes do not affect training, selection, thresholds, or arbitration, and comparator policies are not replayed. The evidence therefore concerns packet-level command response under the tested configuration rather than end-to-end FAROS superiority.

Table~\ref{tab:ns3_actions} reports coherent action responses. Relative to matched $A_0$, resource-increasing $A_1$ and $A_4$ have higher mean goodput and PDR and lower mean delay; the reported 95\% intervals for their goodput and mean-delay differences exclude zero.

\begin{table*}[!t]
\centering
\caption{Independent packet-level ns-3 check of the six frozen commands. Values are means over 20 matched \texttt{RngRun} values per action; final columns give paired 95\% confidence intervals relative to $A_0$.}
\label{tab:ns3_actions}
\resizebox{\textwidth}{!}{%
\begin{tabular}{llccccc}
\toprule
\textbf{Action} & \textbf{Command} &
\textbf{Goodput (Mbps)} &
\textbf{PDR} &
\textbf{Mean delay (ms)} &
\textbf{$\Delta$ Goodput vs.\ $A_0$ [95\% CI]} &
\textbf{$\Delta$ Mean delay vs.\ $A_0$ [95\% CI] (ms)} \\
\midrule
$A_0$ & Keep configuration
& 5.6113 & 0.9300 & 86.41
& -- & -- \\
$A_1$ & Increase PRBs 10\%
& 5.6697 & 0.9377 & 59.46
& $+0.0584$ [$+0.0261,+0.0907$]
& $-26.95$ [$-43.02,-10.87$] \\
$A_2$ & Decrease PRBs 10\%
& 5.5810 & 0.9284 & 119.09
& $-0.0303$ [$-0.0528,-0.0078$]
& $+32.68$ [$+16.59,+48.76$] \\
$A_3$ & Switch scheduler
& 5.5985 & 0.9272 & 84.06
& $-0.0128$ [$-0.0245,-0.0010$]
& $-2.35$ [$-6.59,+1.89$] \\
$A_4$ & Switch scheduler + increase PRBs
& 5.6538 & 0.9348 & 55.47
& $+0.0425$ [$+0.0117,+0.0734$]
& $-30.95$ [$-51.13,-10.76$] \\
$A_5$ & Switch scheduler + decrease PRBs
& 5.5730 & 0.9266 & 127.63
& $-0.0382$ [$-0.0617,-0.0147$]
& $+41.22$ [$+17.75,+64.68$] \\
\bottomrule
\end{tabular}%
}
\end{table*}

Resource-reducing $A_2$ and $A_5$ move packet-level outcomes in the opposite direction, lowering goodput and increasing delay; $A_5$ mean-delay differences are unfavorable in all 20 matched runs. Scheduler-only $A_3$ has smaller mixed effects. Its goodput difference is $-0.01277$ Mbps with 95\% CI [$-0.02453$, $-0.00101$]; the paired $t$-test gives $p=0.03479$, unchanged after Holm adjustment, while Wilcoxon gives $p=0.01531$ unadjusted and $p=0.02189$ after Holm adjustment. Mean/P95 delay intervals cross zero, and there is no Holm-supported P99 improvement.

\begin{figure}[!t]
    \centering
    \includegraphics[width=\columnwidth]{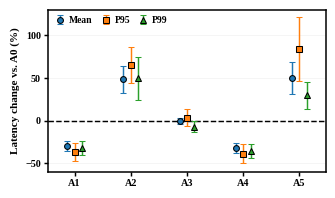}
    \caption{Normalized ns-3 latency response for $A_1$--$A_5$ relative to matched $A_0$; markers show mean, P95, and P99 latency changes, and error bars are 95\% confidence intervals over the same 20 matched runs.}
    \label{fig:ns3_paired_response}
\end{figure}

Fig.~\ref{fig:ns3_paired_response} expresses the matched latency effects on a common relative scale,
\begin{equation*}
\Delta_{\mathrm{rel}}(A_i,m)
=
100\,
\frac{\overline{m(A_i)-m(A_0)}}{\overline{m(A_0)}},
\end{equation*}
where $m$ denotes mean, P95, or P99 delay. This normalization leaves the paired samples, confidence intervals, and statistical tests unchanged. Resource-increasing actions reduce delay, resource-reducing actions increase it, and $A_3$ remains comparatively small and unresolved for mean/P95 delay.

A natural-frequency consistency check weights the action-level responses by the frozen $M_5$ action frequencies. The goodput difference is +0.01012 Mbps with 95\% CI [$-0.00127$, +0.02151], with Holm-adjusted paired-$t$/Wilcoxon $p=0.464/0.715$. Median delay is the narrow exception at about $-0.101$ ms with 95\% CI [$-0.1365$, $-0.0655$]. The broader mixture evidence does not establish packet-level superiority; ns-3 supports only the operational meaning of the commands under matched randomness.

\section{Discussion}
\label{sec:discussion}

\textbf{Assumptions and applicability.} FAROS is evaluated under the frozen trace-conditioned transition--reward contract used for state evolution and A5v2 scoring. The public dataset twins commercial LTE traffic on an experimental Open RAN platform rather than a live deployment, and offered-load tests cover $0.25\times$--$1.25\times$. The independent ns-3 study checks the packet-level semantics of the six commands and does not evaluate closed-loop sequential $M_5$ behavior. Fig.~\ref{fig:oran_architecture} therefore represents a logical deployment architecture. E2SM-KPM, E2SM-RC, A1/O1 signaling, and end-to-end RIC/transport latency are outside the implemented scope.

\textbf{Threats to validity.} The zero executed prohibited-action rate follows from predicate enforcement and does not constitute evidence of learned or formally verified safety. Because the evaluator has no explicit shared PRB budget or inter-cluster contention, the experiments do not isolate a CTDE advantage over independent PPO. The model-matched deterministic rule is a diagnostic reference, while Fair MAPPO uses a different training budget. PCMCI+ provides training-only lagged conditional-dependence support for the temporal-risk path. It does not establish action-level causal effects or determine the historical risk weights. SHAP explains the pre-arbitration actor, whereas final executed commands are covered by the proposal--selection--execution audit.

Reproducibility is supported by the fixed evaluation dataset, MAPPO configurations, matched starts, ten checkpoint hashes, the 150,000-decision trace, analysis outputs, and the ns-3 package. A clean ten-seed evaluation reproduces the reported $M_5$ metrics. Seven late training starts cannot reach validation or final-test states. Incomplete upstream clustering provenance and the unavailable original LSTM training lineage limit reconstruction of those historical stages; accordingly, historical full-trace forecast statistics are not treated as holdout evidence, and the separate leakage-safe experiment provides the disjoint-target forecast test.

\section{Conclusion and Future Work}

FAROS connects QoS forecasting and temporal risk to MAPPO under CTDE and governs execution through symbolic admissibility and reliability-aware Top-$K$ arbitration. The independent $H=8$ test reaches MAE 0.014279 and $R^2=0.961723$. At $H=3$, temporal risk raises vulnerability recall from 0.083592 to 0.325462 and balanced accuracy from 0.540750 to 0.653257. The same-actor shield removes the 6.67\% risky-state prohibited-action rate, while $M_5$ reaches reward 0.27781 and recovery 90.56\%. The model-matched rule remains higher at 0.30151 and 97.57\%, so no superiority over direct threshold logic is claimed. A5v2 changes 22.61\% of proposals, while the zero executed prohibited-action rate is enforced by the predicates. The 120 ns-3 executions provide action-level evidence only. Future work should use state-synchronized closed-loop O-RAN/ns-3 evaluation and recalibrate temporal risk under substantial distribution shift.

\bibliographystyle{IEEEtran}
\bibliography{References}

\end{document}